\documentclass{cas-sc}

\usepackage{amssymb}
\usepackage[authoryear,longnamesfirst]{natbib}

\usepackage{rotating}
\usepackage{tabularx, booktabs, makecell, caption}
\usepackage{adjustbox} 
\usepackage{textgreek}
\usepackage{array}
\usepackage[gen]{eurosym}
\usepackage{cleveref}
\usepackage{footnote}
\usepackage{longtable}

\makesavenoteenv{tabular}
\usepackage[flushleft]{threeparttable}
\usepackage{booktabs}
\usepackage{float}
\usepackage{natbib}
\usepackage{chngpage}
\usepackage[section]{placeins}
\usepackage{pdfpages}
\graphicspath{{Figs/}} 
\newcommand\xw{1}

\usepackage{xcolor}
\definecolor{dgreen}{rgb}{0.15, 0.80, 0.10}

\def\tsc#1{\csdef{#1}{\textsc{\lowercase{#1}}\xspace}}
\tsc{WGM}
\tsc{QE}

\begin{document}
\let\WriteBookmarks\relax
\def\floatpagepagefraction{1}
\def\textpagefraction{.001}


\title[mode = title]{Effects of seat back height and posture on 3D vibration transmission to pelvis, trunk and head}

\shorttitle{Seat back and posture}    
\shortauthors{Mirakhorlo et al.}

\author[1]{Mojtaba Mirakhorlo}
\cormark[1]
\fnmark[1]

\author[2]{Nick Kluft}

\author[1]{Barys Shyrokau}

\author[1]{Riender Happee}

\affiliation[1]{organization={Cognitive Robotics},
            addressline={Mekelweg 2}, 
            city={Delft},
            postcode={2628 CD}, 
            state={Zuid-Holland},
            country={The Netherlands}}

\affiliation[2]{organization={Department of Human Movement Sciences, Vrije Universiteit Amsterdam},
            addressline={van der Boechorststraat 7-9},
            postcode={1081 BT Amsterdam}, 
            state={Noord-Holland},
            country={The Netherlands}}

\cortext[1]{Corresponding author}

\begin{abstract}
Vibration transmission is essential in the design of comfortable
vehicle seats but knowledge is lacking on 3D trunk and head motion
and the role of seat back and posture. We hypothesized that head
motion is reduced when participants' upper back is unsupported, as
this stimulates active postural control. We developed an experimental
methodology to evaluate 3D vibration transmission from compliant
seats to the human body. Wide-band (0.1-12 Hz) motion stimuli were
applied in fore-aft, lateral and vertical direction to evaluate the
translational and rotational body response in pelvis, trunk and head.
A standard car seat was equipped with a configurable and compliant
back support to test 3 support heights and 3 sitting postures (erect,
slouched, and preferred) where we also tested head down looking at a
smartphone. 

\noindent Seat back support height and sitting posture substantially affected vibration transmission and affected low frequency
responses in particular for body segment rotation. According to our
hypothesis a low support height proved beneficial in reducing head
motion. 

\noindent \textit{Relevance to industry:} Our methodology effectively evaluates 3D
wide-band vibration transmission from compliant seats to the human
body. The lowest back support height reduced head motion but was
perceived as least comfortable. This calls for seat designs which
support but do not so much constrain the upper back. The head down
posture enlarged head motion, pleading for computer system
integration allowing heads up postures in future automated cars. The
biomechanical data will serve to validate human models supporting the
design of comfortable (automated) vehicles.
 
\end{abstract}




\begin{keywords}
\sep Vibration \sep Comfort \sep Posture \sep Automotive \sep Seat
\end{keywords}

\maketitle


\section{Introduction}
\label{sechh}
Vibrations transmitted from the road to the human body through the seat affect perceived motion comfort  \citep{corbridge1986vibration,tiemessen2007overview,dong2019effect}. Particularly in automated vehicles, motion comfort is essential \citep{kyriakidis2015public} as these vehicles are intended for spending time on work and leisure activities. Assessment of postural stabilization and comfort can be used in seat design \citep{papaioannou2021assessment}
 and in motion planning \citep{zheng2021comfort} of automated vehicles, potentially resulting in higher comfort levels. 

The human response to seat vibration is usually quantified by measuring the seat-to-head transmissibility (STHT) including resonance frequencies which are assumed to relate to the level of discomfort \citep{paddan1998review,rahmatalla2010quasi}. Human postural responses to vibrations have been studied by investigating the effect of seat configuration factors on STHT, such as back support inclination \citep{basri2014application,jalil2007fore,Nawayseh201582}, seat pan inclination \citep{jalil2007fore}, and thickness of foam cushions \citep{zhang2015transmission}. Back support height, on the other hand, has been investigated only in few studies. \cite{toward2011transmission,toward2009apparent} compared vertical loading with and without back support and reported higher resonance frequencies with back support. 
\cite{Jalil2007} showed that back support height hardly affected the resonance frequency of the back support in the fore-aft direction but did not report human body responses. 

Human sitting posture, in combination with the configuration of the car seat, can affect postural stability, where the seat may promote postures that enhance trunk stability and comfort. \cite{bhiwapurkar2016effects} and \cite{song2017subjective} studied effects of posture sitting without back support. \cite{bhiwapurkar2016effects} showed that forward-leaning sitting postures cause an extra peak in STHT in comparison with erect sitting postures. Mansfield \citeyearpar{mansfield2006effect} studied the apparent mass in ``relaxed'' and ``tense'' sitting postures and report a stronger non-linearity in relaxed postures. \cite{ADAM2020103014} showed significant effects of posture and backrest usage on vertical transmission in a suspended rigid seat but did not report body motion and perceived comfort. However, we are not aware of studies on the effect of erect versus slouched sitting postures on 3D body kinematics, underlying postural control strategies and perceived comfort in car seats. 



For a better understanding of biomechanical responses to vibration, relevant body segment responses in the mechanical chain between head and seat (pelvis and trunk) need to be considered. The pelvis response is essential for cushion design, and the trunk response for back support design. Most studies have focused only on head responses, i.e STHT in either vertical   \citep{boileau1998whole,fairley1989apparent,toward2011transmission}, fore-aft   \citep{nawayseh2005non,nawayseh2020tri} or lateral perturbations \citep{mandapuram2012apparent,bhiwapurkar2016effects}.  
Many studies investigated seat-to-head transmissibility on rigid seats, which reduces complexity. However, rigid seats are inherently uncomfortable \citep{Li2020} and thereby less suited to study effects of posture and support on perceived comfort. Finally, most studies only report 1D results such as the head vertical response to seat vertical motion. Hence, these studies neglect secondary motion reactions in other translational directions, and ignore body segment rotations which will also affect perceived comfort \citep{paddan1998review}.
Several studies report the apparent mass at seat and back support but do not study body and head motion \citep{mansfield2006effect,JunWuRigid,JunWuTrain}.
\par In the current study, we investigate how sitting posture and seat back height affect the perceived motion comfort and the transmissibility of motion between the car seat and the human body. We jointly analyze pelvis, trunk, and head 3D translation and rotation in responses to fore-aft, lateral and vertical seat motion. In order to achieve realistic seat interaction and comfort levels, we use a commercial car seat pan and a simple but compliant seat back support. 
We hypothesize that postural stabilization and comfort will depend on back support height. In studies with unsupported back applying fore-aft platform motion we found that participants could effectively minimize head rotations in space \citep{van2016trunk} while with rigid full back support and harness belt substantial head rotations emerged \citep{forbes2013dependency}. We hypothesize that a full back support constrains lumbar and thoracic spine motion and prevents a coordinated full spine control strategy. Hence we expect amplified head rotation with full back support. Conversely we expect a low backrest to be beneficial for head stabilization, as it promotes the back to actively stabilize the trunk and head. 

We study effects of back support height and posture including slouched, preferred, and erect postures, as well as a head-down posture representing usage of digital devices. This head-down posture might become very common in automated vehicles as they allow occupants to work on a tablet, laptop, or smartphone without being a hazard to other road users. Furthermore, we vary motion amplitude to quantify the non-linearity of postural responses \citep{mansfield2006effect}. For future modeling of postural stabilization, we evaluate the influence of vision on postural control where we expect small but significant effects of vision on trunk \citep{van2016trunk} and head \citep{forbes2013dependency} stabilization.  

To achieve these scientific objectives we developed an experimental methodology to evaluate 3D vibration transmission from compliant seats to the human body. We designed wide-band motion stimuli and applied these in fore-aft, lateral and vertical direction and evaluated the translational and rotational body response in pelvis, trunk and head, and used analyses of variance to assess significance of the effects of posture and seat back height across seat motion directions.

\section{Methods}

Eighteen healthy adults (9 male, 9 female) participated in this study. Participants were balanced on age group (30-39, 40-49, and 50-60 years) and sex (for participant descriptives see Table A.1 in Appendix \ref{sec:sample:descriptives}). Inclusion criteria were that participants considered themselves healthy. 
Prior to any experimentation, participants were informed on the procedures and goals of the experiment by an information letter, and once again just before the start of the experiment. Participants provided written informed consent. The study protocol was approved by the Human Research Ethics Committee of the Delft University of Technology (HREC 962). During the experiment, participants were closely monitored on their well-being and we evaluated their misery after each trial (a long break was offered when MISC\textgreater 4). All participants were reimbursed with a \EUR{20} gift card.

Participants were instructed to sit in a car mock-up, mounted on top of a six-degrees-of-freedom motion platform \citep{khusro2020mpc}. The mock-up consists of the cockpit of a Toyota Yaris and participants were seated in the modified passenger’s seat (see Figure 1).

\begin{figure}
\centering
\includegraphics[width=0.8\textwidth]{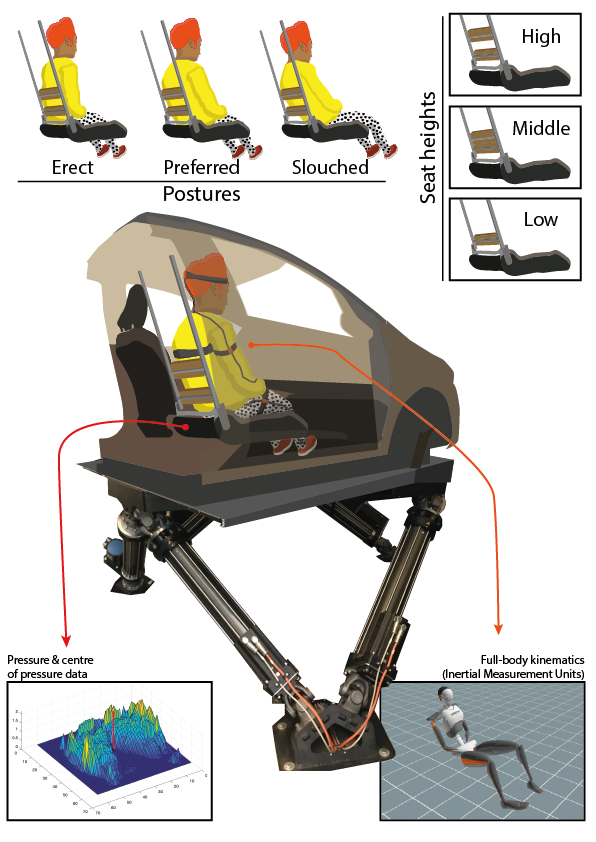}
\caption{Experimental setup. A front part of a car was mounted on top of a six-degrees-of-freedom motion platform. A bottom part of a Toyota Yaris passenger seat was used for this experiment. The original back support was replaced by a steel frame that is stiff and allowed for easy adjustment of the back support height. The frame was equipped with either one or two cushion pads that were in direct contact with the back: a low support pad and a high support pad. The pitch angle of the upper cushion pad could be adjusted to better match the shape of the back of the participant. A pressure mat was mounted on top of the seat to record the pressure distribution between the participant and the seat. \label{fig:set_up}}

\end{figure}

\subsection{Input vibrations}
\begin{figure}[h!]
\centering
\includegraphics[width=0.9\textwidth]{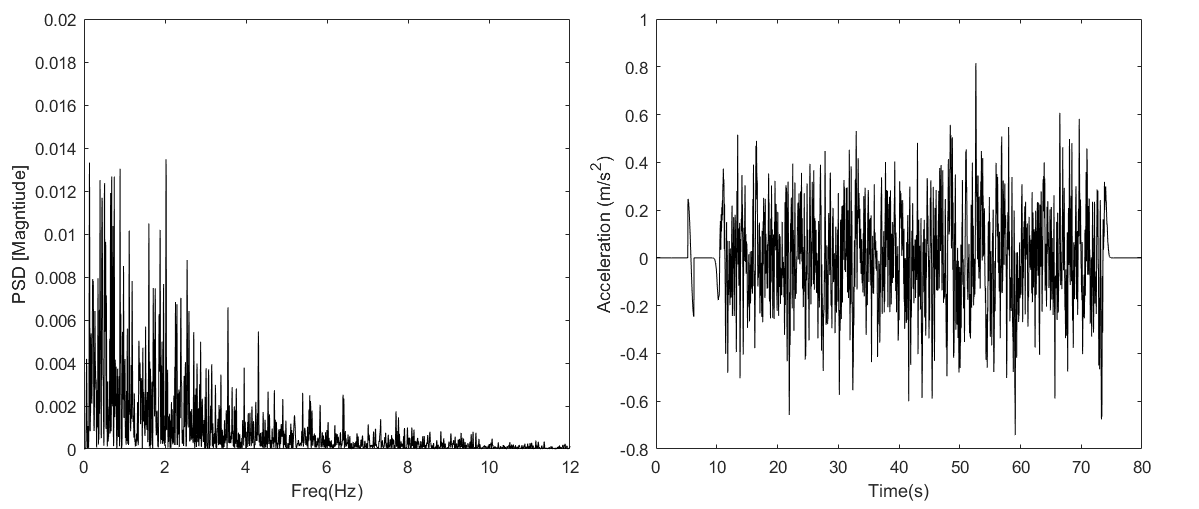}
\caption{Power spectral density of the platform’s input signal (left panel) and part of the input signal in time domain (right panel) applied in fore-aft direction. Excitation signals for lateral and vertical directions are exactly the same as fore-aft. Power is greatest in the 0.1-3 Hz frequency band, between 3-12 Hz power is reduced but still significant. Hardly any power is present above 12 Hz. The pulse at t=6 s served for time synchronization.}
\label{fig:inputsignal}
\end{figure}

A wide-band noise signal was designed as input for the motion platform (see Figure \ref{fig:inputsignal}). Similar signals were used to study the human response to bicycle vibrations \citep{dialynas2019}. We adapted the signal to be more comfortable and to better approximate car driving while maintaining a good coherence between the applied platform motion and the measured body response at the pelvis, trunk, and head and fitting within the working envelope of the  motion platform. In short, the signals comprised random noise with a frequency bandwidth of 0.1 - 12.0 Hz and 0.3 m/s\textsuperscript{2} rms power. This range was selected to include low/mid frequencies with postural stabilization using visual, vestibular, and muscle feedback and high frequencies dominated by the passive body and seat properties including resonance frequencies around 4-6 Hz in vertical loading. This range also includes low frequencies associated with motion sickness causation but duration and amplitude of the 12 trials were limited aiming to prevent actual motion sickness to develop as this would invalidate following trials. The resulting vertical motion resembles driving at somewhat uncomfortable roads. Horizontal vehicle motions will have less power at mid and higher frequencies but we chose to apply the same motion as in vertical as a lower amplitude would preclude the attainment of coherent results using frequency domain analysis. 

For each seat back and posture condition, one trial was performed. Each trial lasted 200 s and sequentially applied motions in three different axis directions (i.e., fore-aft, lateral, vertical) with 60-second duration, 3-second fade-in, and 3-second fade-out to avoid abrupt motions.   

\subsection{Postures \& back support height}
Participants were subjected to the vibrations in three main posture conditions: sitting erect, slouched, and sitting in the subject’s preferred posture. The erect posture aimed to achieve an S-shaped back curvature. We instructed participants to sit straight such that their belly was pressed out while their back made a hollow arch, with their buttocks at the most posterior position of the seat.
To obtain a slouched posture, participants were instructed to move the pelvis forward while keeping their chest straight, which flexes the lumbar spine towards a C shape curvature. 
Prior to the first slouched condition, we instructed the participants to sit in the middle of the seat with their belly pressed inward, trying to bend the lower back as much as possible, while keeping the upper chest straight. 
We verified the posture by evaluating the pitch angle of the thorax with respect to the pelvis as measured in real-time using the motion capture system (see below), and if needed we gave feedback to the participants. Post hoc analysis of the angle between thorax and pelvis relative to the horizontal showed that the angle between the horizontal and the line connecting the pelvis Centre of Mass (CoM) with the thorax CoM was largest for the erect posture (mean angle = 65.1\textdegree), followed by the preferred posture (mean angle = 62.4\textdegree). The angle was sharpest when participants were slouching (mean angle = 53.6\textdegree, see Figure \ref{sec:sample:conditions} in Appendix A for more detail on postural differences between conditions).

Besides the sitting posture, the back support height was varied (see Fig. \ref{fig:set_up} for an overview of the different conditions). The seat was equipped with a lower and an upper back support pad (both 11.6 cm height x 42.0 cm width x 6 cm thick cushion pads mounted on flat plates). 
The lower back support pad was fixed to the chair, while the upper back support pad could be taken off and could be shifted vertically, along two beams that were 20 degrees backward rotated (i.e., backrest made a 110-degree angle with the horizontal). The pad could be re-oriented in pitch to match the curvature of the subject’s back. Participants were subjected to three back support conditions: 1) low back support (lower pad only; the top of the support pad was situated at the height of the posterior superior iliac spine), 2) mid back support (two pads; the upper back support pad was placed on top of the lower support pad, thereby covering the pelvis and lumbar spine), and 3) high support (two pads; also supporting the thorax - the top of the upper support pad was aligned with the apex of the scapula's angulus inferior, which was identified by palpation). Changing the back support height might induce an undesired change in sitting posture; however, a post hoc analysis on the effect of back support height showed only very subtle changes. 

\subsection{Measurement protocol}
Participants were guided on top of the platform and took place in the experimental seat, mounted in the passenger’s position of the car mock-up. The seat belt was not fastened as modern belts exert marginal forces in normal driving and the belt might interfere with instrumentation and cables. Participants looked straight ahead through the windshield of the car mock-up (Figure \ref{fig:set_up}). 

\begin{figure}
    \centering
    \includegraphics[width=\textwidth]{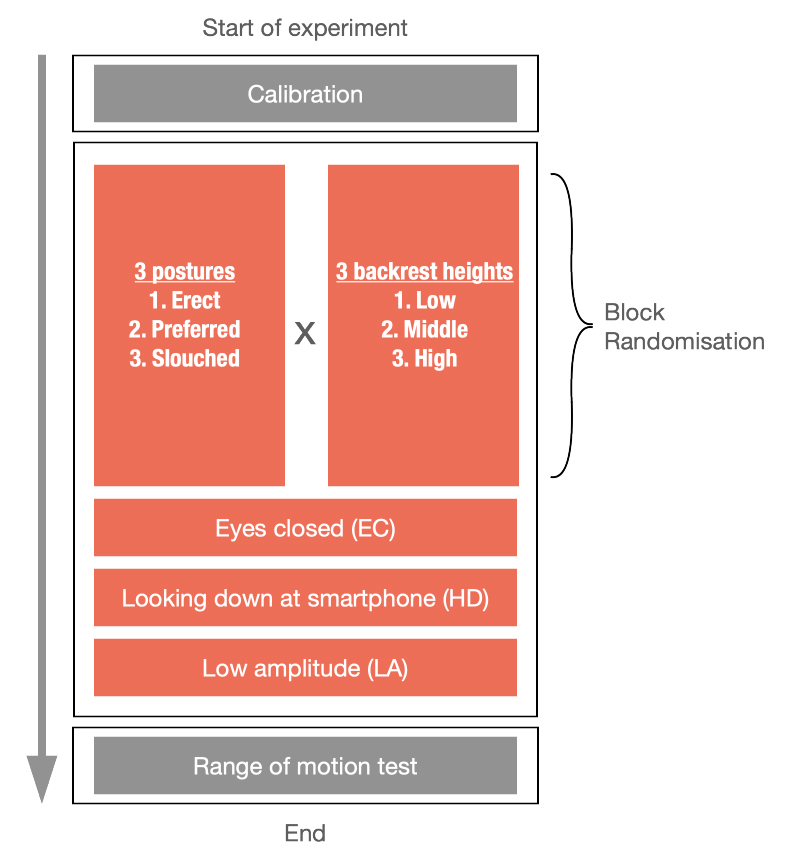}
    \caption{Schematic overview of the experimental protocol. An experiment began with a set of calibration postures to estimate the position of the body segments. Next, participants were subjected to 9 trials combining three postures with three back support heights. During these trials participants had their hands folded on their lap and gazed straight ahead through the windshield. Three additional trials followed where the participants had their eyes closed (EC), head down looking at a smartphone (HD), or with low amplitude input vibration being 25\% of the original amplitude (LA).}
    \label{fig:scheme}
\end{figure}

Participants were subjected to 12 conditions shown in Figure \ref{fig:scheme}. Participants were allowed to take breaks between conditions to prevent drowsiness and discomfort due to prolonged sitting. We tested 3 postures for 3 seat back configurations as described above. In three additional trials, subjects sat in their preferred posture, with middle-back support. They were  subjected to the same platform vibrations, but now with respectively 1) their eyes closed (EC), 2) looking down at a turned-off smartphone (i.e., head down, HD), or 3) looking forward with a lower input vibration amplitude (0.25 times the original amplitude, LA). The order of conditions was randomized across participants. However, adjusting the back support height was time-consuming. Therefore the randomisation was performed at two levels, firstly randomizing the order of the three back support heights, and secondly randomizing the other variations within each back support block.

Finally the active ranges of motion of the entire spine (flexion/extension and lateral flexion) were recorded while standing using a protocol from \cite{frey2020} to support future modelling (see Appendix A).

\subsection{Measurement devices}
\subsubsection{Perceived discomfort \& sway}
Perceived discomfort and perceived sway were assessed using a 9-item questionnaire. This consisted of three main parts assessing 1) subjective Misery using the misery scale (MISC) \citep{Reason1975}, 2) perceived discomfort \citep{vanVeen2015}, and 3) perceived sway (modified from \cite{vanVeen2015}). Besides the overall discomfort of the chair and backrest, seven questions concerned the perceived discomfort and sway specific for the studied body parts (trunk, lower and upper back, and neck). Participants filled the questionnaire for each individual excitation direction. The full questionnaire can be viewed on the experimental data repository. 
Prior to any experimentation, participants were given some time to study the items of the questionnaire. After each trial, the experimenter read the questionnaire's items out loud, and participants verbally responded by rating their misery on a 0-9 scale, and their perceived discomfort and perceived sway on a 1-10 scale (i.e., a high score corresponded to high discomfort or sway and vice versa). Perceived sway was mentioned by the first participants and formally reported and analyzed starting from the fifth participant (N=14). 

\subsubsection{Kinematics}
The platform acceleration was recorded by three triaxial acceleration sensors mounted at the upper part of the motion platform, and equally distributed on a circle with a radius of 0.5 meters, with a sampling frequency of 100 Hz.
To capture the 3D whole-body kinematics, participants wore a motion capture suit with seventeen triaxial inertial measurement units at 240 Hz (MTW Awinda, Xsens Technologies, Enschede, The Netherlands). Before experimentation, for each participant, calibration postures were recorded and circumferences of body segments were measured using measurement tape. Through integration, the Xsens software reconstructs orientations of all body segments and the quasi-global positioning of their joints \citep{white2018}. On the basis of the reconstructed joint positions, the center of mass (CoM) of each body segment was estimated \citep{Zatsiorsky2002}. Segment accelerations were projected on these estimated CoM positions. In this paper, we use the reconstructed 3D motion of the pelvis, trunk, and head center of gravity, presented in world coordinates (X=forward, Y=left, Z=up).

\subsubsection{Seat pressure \& electromyography}
Seat pressure was recorded using the XSENSOR X3 medical seat system (XSensor Technology Corporation, Calgary, AB, Canada). Pressure was recorded at the buttocks and thighs, with a grid of 48 $\times$ 48 sensors at a 1.27 cm distance between load cells and used to estimate the center of pressure and the resultant force (see Appendix B).

Muscular activity (EMG) recorded in 4 participants showed a disappointing coherence to the applied motion stimuli in particular for lumbar muscles, and was therefore not recorded in other subjects (Appendix C).

\subsection{Data analysis}
Data was synchronized using recorded timestamps for kinematics, seat pressure and EMG while platform acceleration was synchronized using a pulse applied at the onset of platform motion (Figure \ref{fig:inputsignal}).

\subsubsection{Perceived comfort \& sway} 
To quantify the overall discomfort, and the perceived trunk and head sway within conditions, the ratings of the overall seat-discomfort and perceived-sway items were averaged over the 3 excitation directions. 

\subsubsection{Kinematics\label{text:freqresponse}} 
Platform accelerations were up-sampled to 240 samples per second to match the body kinematic data. 
To evaluate the transmission from platform motion to body segment motion, transfer functions were calculated, for each condition for each individual participant. The Hanning function was used for windowing the signal in 15 segments (i.e., a window size of 24 seconds) with 50 percent overlap. Gain, phase, and coherence were calculated for the linear (fore-aft, lateral, and vertical) and rotational accelerations (roll, pitch, and yaw) of body segments (pelvis, trunk, and head) in response to the measured platform accelerations.  
To quantify the effect of seat back support and sitting posture, peaks and related frequencies of response gains were analyzed. Peaks were analyzed for the main translational responses (fore-aft to fore-aft, lateral to lateral, and vertical to vertical), and main rotational responses (pitch to fore-aft, roll for lateral, and pitch for vertical). The peak search was constrained to frequencies where consistent peaks were observed across participants. Regarding the fore-aft responses to fore-aft perturbations, the peak search for pelvis and head was constrained between 2 and 7 Hz. For lateral head responses to lateral perturbations, the peak search was constrained to frequencies below 2.2 Hz. 
The peak search for rotational responses was constrained between 1 and 6.5 Hz. In addition, low-frequency gains were derived for both translational and rotational responses using the average gains between 1 and 2 Hz. 

 
\subsection{Statistical analysis} 
A repeated measures ANOVA was used to statistically test the effects of seat backrest height, and sitting posture. Four factors of direction, body segment, posture, and backrest height have been included in the statistical model. 
A repeated measures ANOVA was also used to statistically test if there are any significant differences between extra trials (eyes closed, head down, and low amplitude) and the corresponding reference trial (middle support and preferred posture). Repeated measures ANOVAs were performed separately for peak translational and rotational gains and their related frequencies, and for low-frequency gains between 1-2 Hz. In case of significant interactions, post hoc comparison tests (i.e., t-tests) were performed with Bonferroni corrections. 
The above statistical analyses were performed after log transformation to enhance normality. For these analyses, Matlab's statistical functions were used.
\par

\section{Results} 
Seventeen participants finished the complete experiment reporting acceptable comfort levels (median MISC = 2, interquartile range = 2). Participant 18 (female, 40-49 years) dropped out as the vibrations led to severe motion sickness (MISC = 8) after 5 out of 12 trials. Participant 6 showed deviant kinematics and was excluded from all kinematic analyses.
\subsection{Perceived comfort \& sway} 

\begin{figure}[h!]
\centering
\includegraphics[width = 1\textwidth]{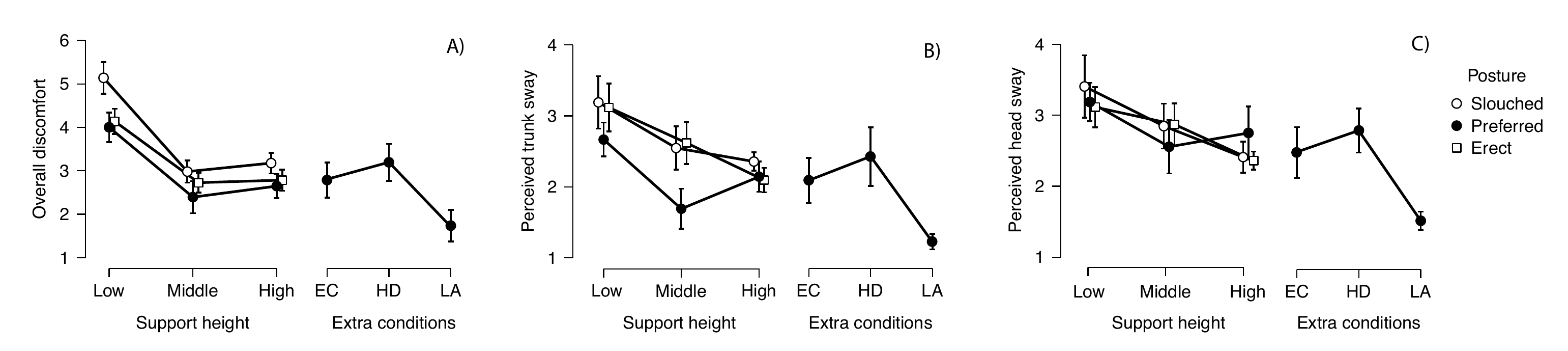}
\caption{Perceived overall discomfort (\textbf{A}), perceived trunk sway (\textbf{B}), and perceived head sway (\textbf{C}). Participant's ratings were averaged over excitation directions. Data points show mean ratings over participants, and error-bars indicate the mean rating $\pm$ standard error. Data are shown for every back support height and posture combination, and the extra conditions} eyes closed (EC), head down (HD), and low amplitude (LA). In the extra conditions participants sat in their preferred posture with middle back support height.\label{perc_support}
\end{figure}

Perceived overall discomfort was modulated by both posture and back support height (posture: F(2) = 10.21, p \textless0.01, support : F(2) = 14.61, p \textless0.001, see Figure  \ref{perc_support} for a graphical overview). A low support was perceived as more discomforting than mid (Cohen's \textdelta{} = 1.19, t = 4.91, p\textsubscript{bonf} \textless0.001), and high (Cohen's \textdelta{} = 1.07, t = 4.42, p\textsubscript{bonf} \textless0.001) back support heights. Similarly, a slouched posture was rated as more discomforting than the preferred (Cohen's \textdelta{} = 1.06, t = 4.37, p\textsubscript{bonf} \textless0.001), and erect postures (Cohen's \textdelta{} = 0.77, t = 3.19, p\textsubscript{bonf} = 0.01). The interaction effect of posture$\times$back support height did not explain the data (F(4) = 0.80, p = 0.53). 
The perceived head sway was not modulated by either posture or back support height (posture: F(2) = 1.77, p = 0.190, backrest: F(2) = 0.14, p = 0.867). A main effect of back support height (F(2) =3.78, p = 0.036) and posture (F(2)= 3.68, p = 0.039) was present on perceived trunk sway. Although not significant, the largest effect sizes in the post hoc comparison were found between low and high back support (Cohen's \textdelta{} = 0.67, t = 2.51, p\textsubscript{bonf} = 0.056), and between slouched and preferred (Cohen's \textdelta{} = 0.68, t = 2.53, p\textsubscript{bonf} = 0.053).

Regarding the additional conditions (eyes closed, head down, low amplitude), the discomfort and perceived sway were affected by condition (overall discomfort: F(3) = 4.16, p = 0.012, perceived trunk sway: F(3) = 6.69, p = 0.008 (corrected for sphericity by Greenhouse-Geisser correction), perceived head sway: F(3) = 6.01, p = 0.002).
Post hoc tests showed that vibrating at a lower amplitude decreased the perceived head sway compared to the reference condition (Cohen's \textdelta{} = 0.94, t = 3.39,  p\textsubscript{bonf} = 0.010 but did not significantly affect perceived discomfort (Cohen's \textdelta{} = 0.50, t = 1.87, p\textsubscript{bonf} = 0.418). The head down and eyes closed conditions did not differ significantly from the reference condition for either overall discomfort or perceived trunk or head sway.


\subsection{Kinematics} 
\begin{figure}[p]
\centering
\includegraphics[width=\xw\textwidth]{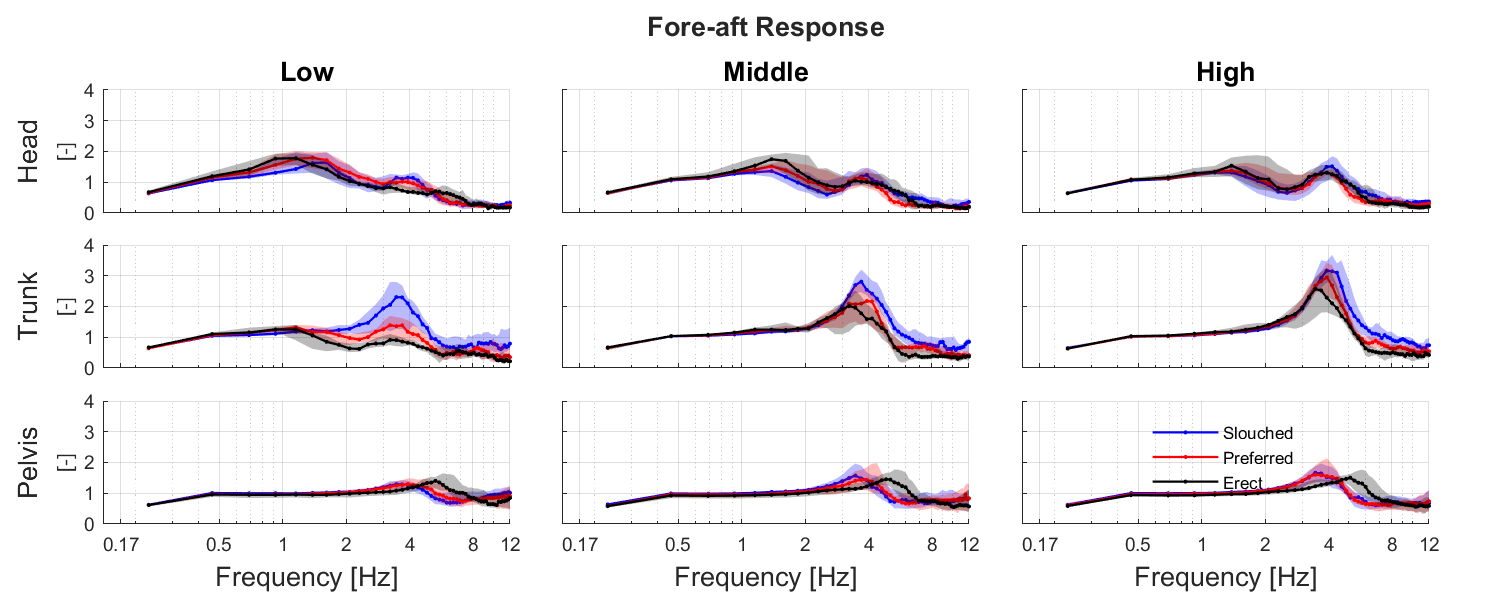}
\includegraphics[width=\xw\textwidth]{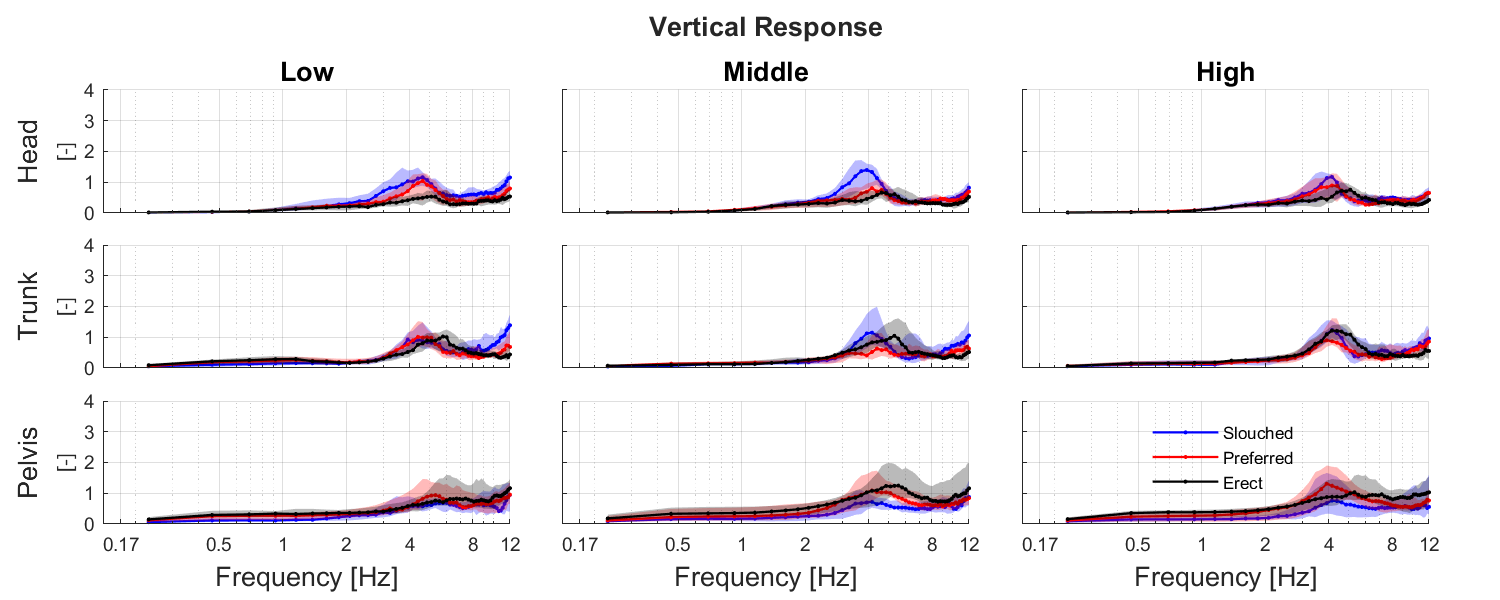}
\includegraphics[width=\xw\textwidth]{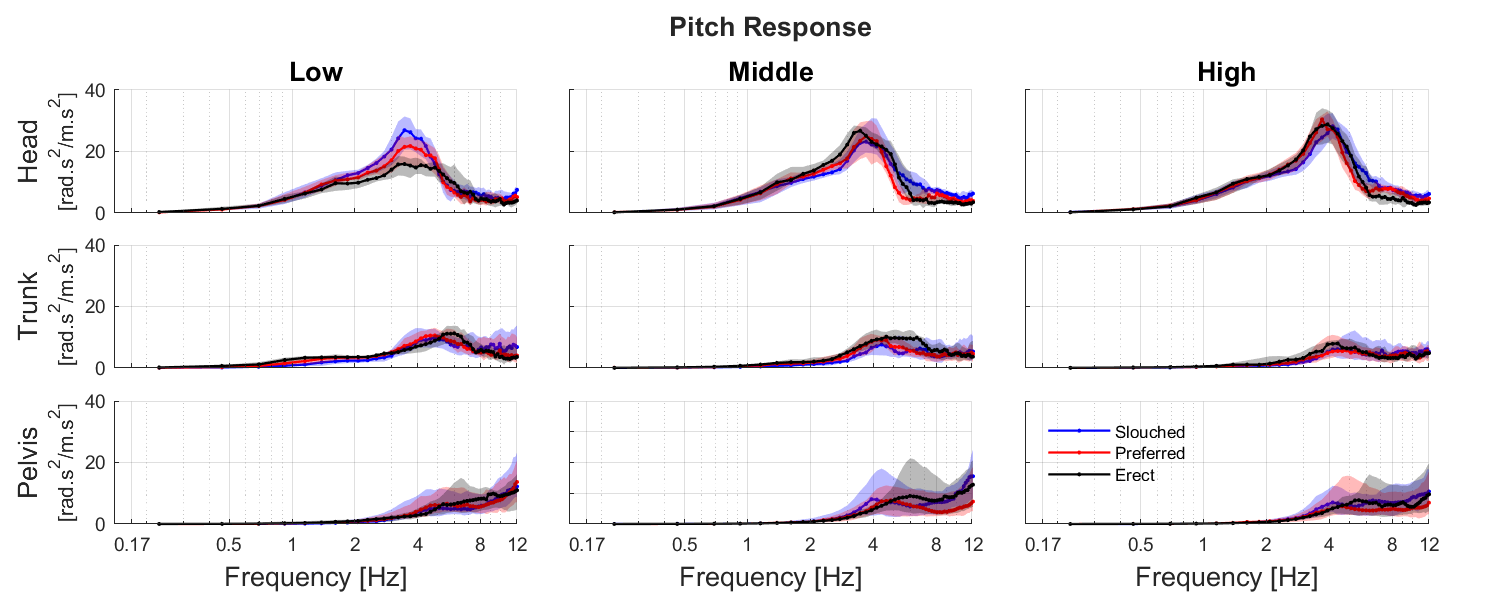}
\caption{Fore-aft perturbations: Fore-aft (top panel), vertical (mid panel) and pitch (lower panel) responses. Median of gains (solid lines) with 25\textsuperscript{th} and 75\textsuperscript{th} percentile (shadows) for low (left), middle (mid) and high (right) back support in slouched, preferred and erect postures. }
\label{Figure:X}
\end{figure}

\begin{figure}[p]
\centering
\includegraphics[width=\xw\textwidth]{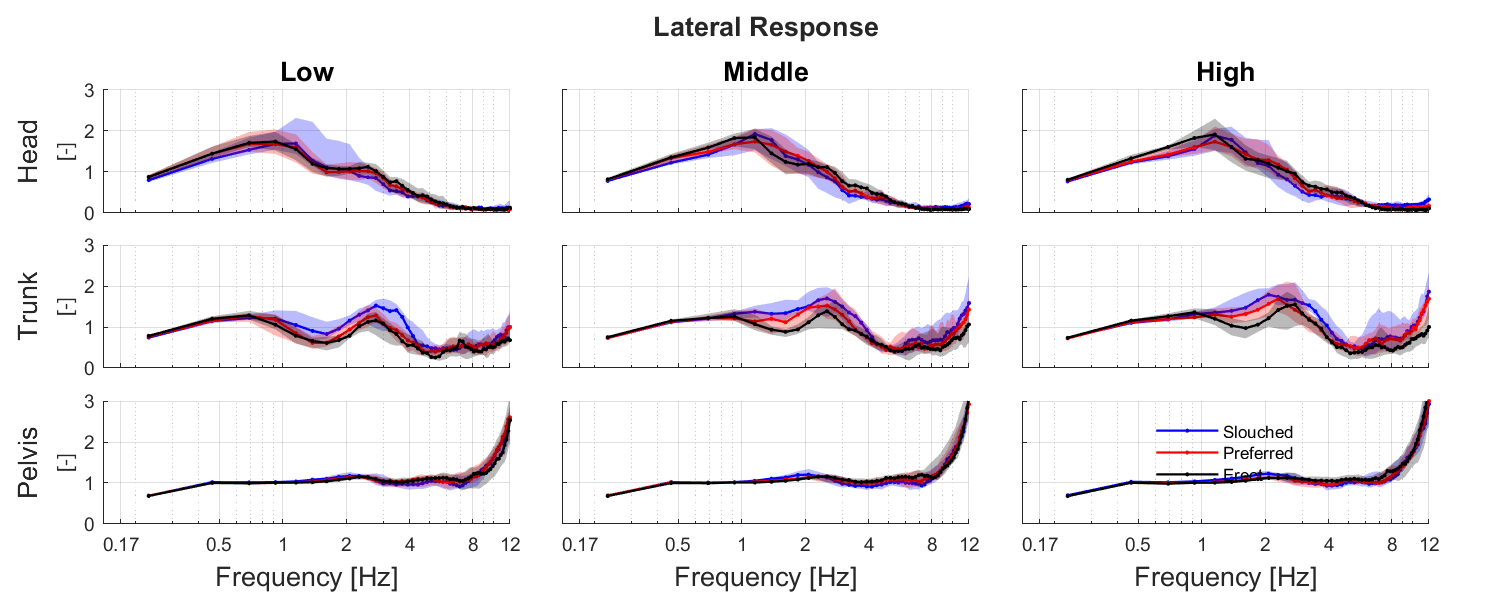}
\includegraphics[width=\xw\textwidth]{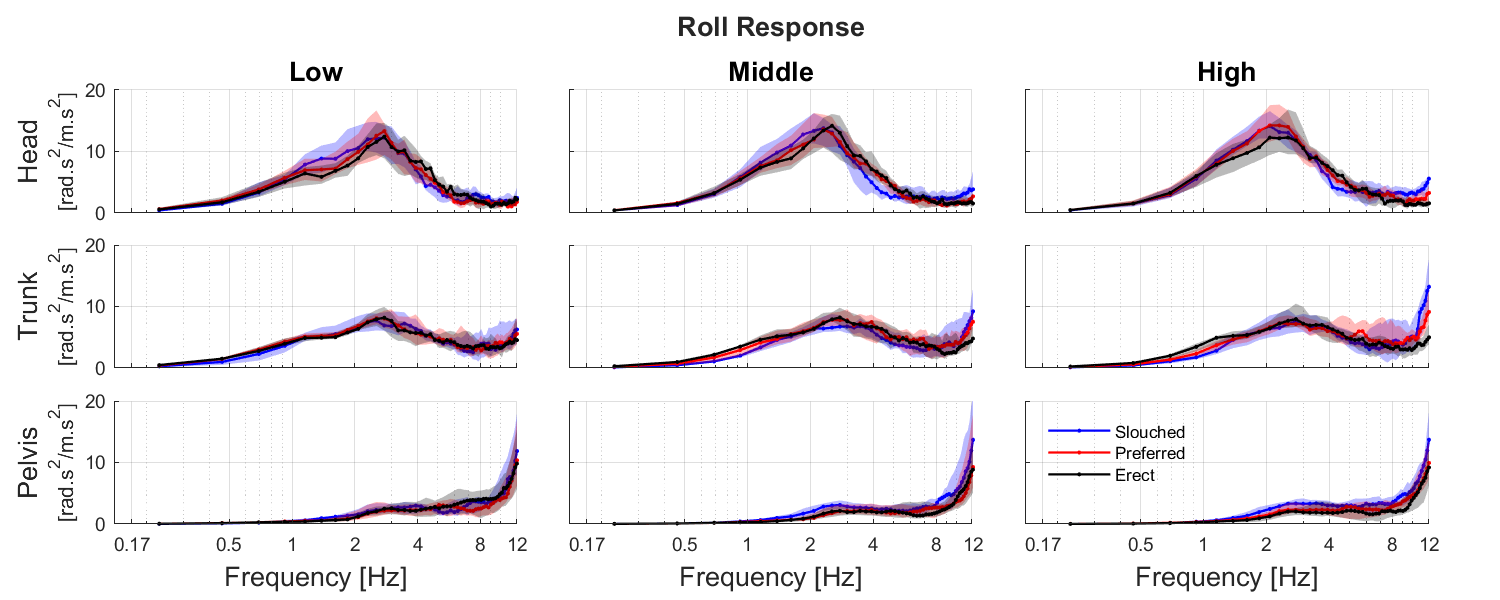}
\includegraphics[width=\xw\textwidth]{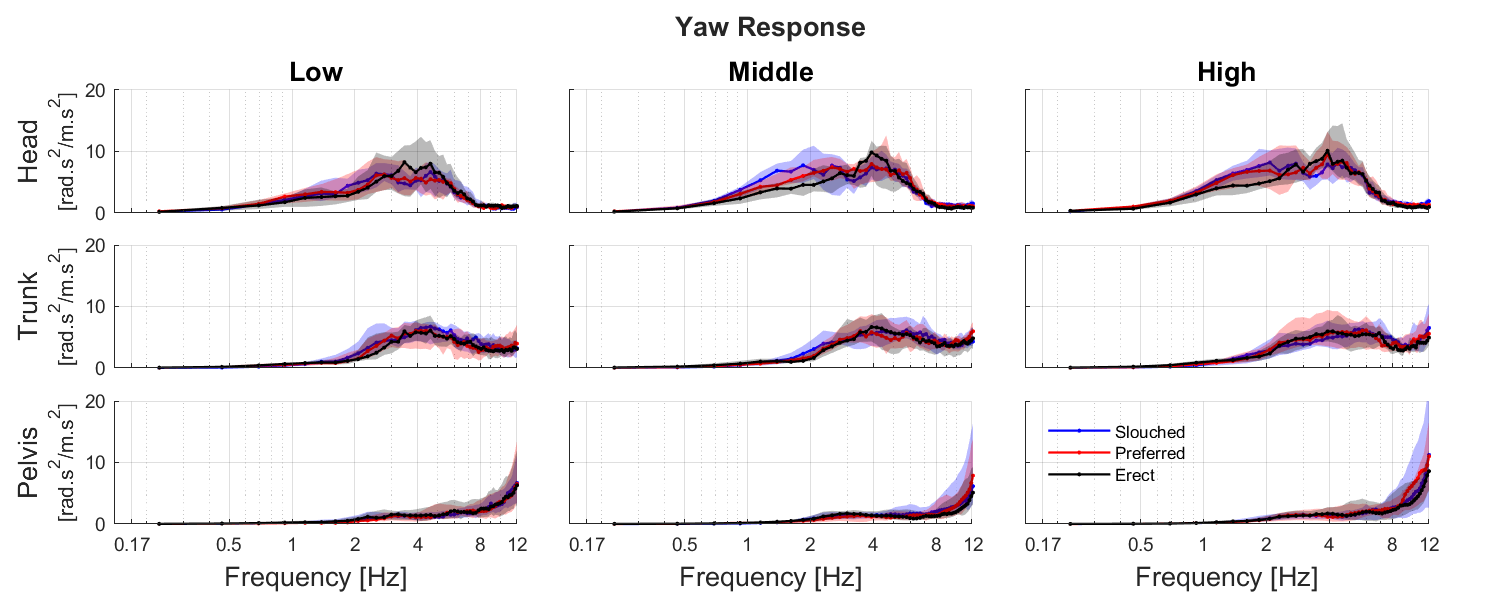}
\caption{Lateral perturbations: Lateral (top panel), roll (mid panel) and yaw (low panel) responses. Median of gains (solid lines) with 25\textsuperscript{th} and 75\textsuperscript{th} percentile (shadows) responses for lateral perturbations for low (left), middle (mid) and high (right) back support in slouched, preferred and erect postures. }
\label{Figure:Y}
\end{figure}

\begin{figure}[p]
\centering
\includegraphics[width=\xw\textwidth]{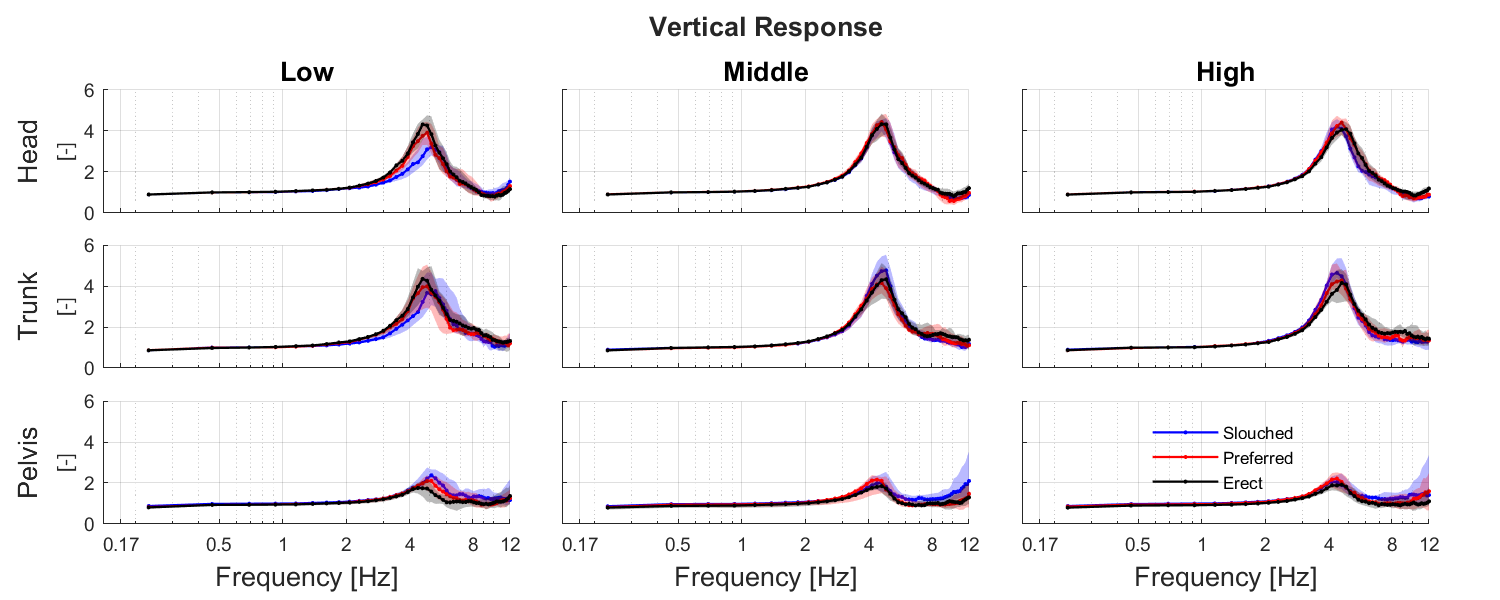}
\includegraphics[width=\xw\textwidth]{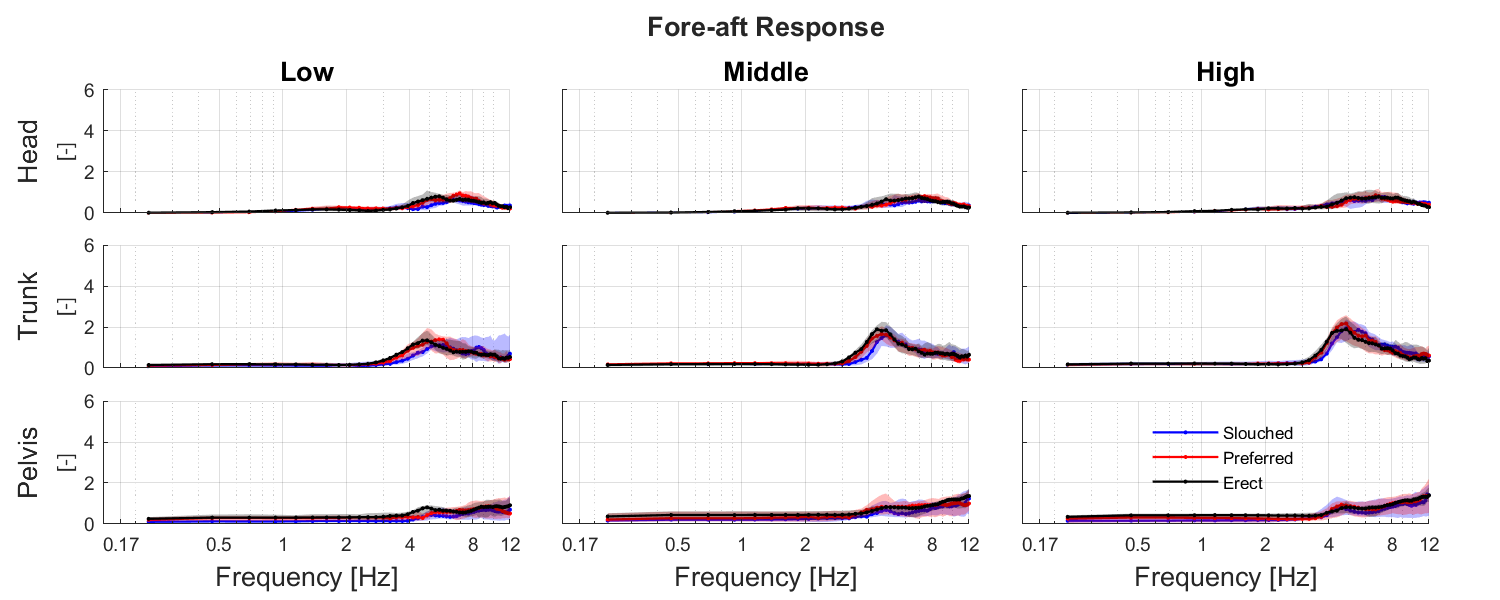}
\includegraphics[width=\xw\textwidth]{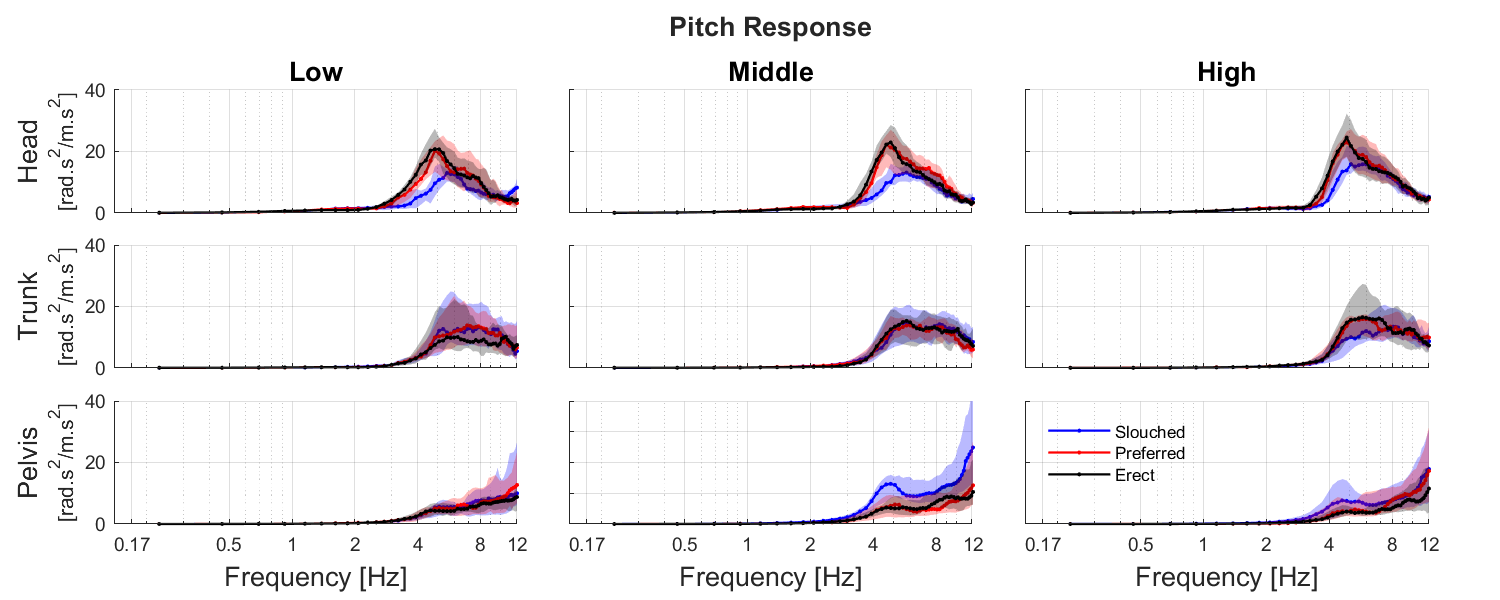}
\caption{Vertical perturbations: Vertical (top panel), Fore-aft (mid panel) and pitch (lower panel) responses. Median of gains (solid lines) with 25\textsuperscript{th} and 75\textsuperscript{th} percentile (shadows) for low (left), middle (mid) and high (right) back support in slouched, preferred and erect postures. }
\label{Figure:Z}
\end{figure}

 \Cref{Figure:X,Figure:Y,Figure:Z} show the effects of backrest height and sitting posture on the head, trunk, and pelvis responses during fore-aft, lateral, and vertical perturbations. These figures show gains for the 3 most relevant responses for each platform motion direction. In all cases, the response in the applied motion direction is shown in the upper section, while the middle and lower sections show interaction terms. For the fore-aft platform motion, \Cref{Figure:X} shows the fore-aft response in the upper panel, the vertical response in the middle panel, and the pitch response in the lower panel. Likewise for the lateral platform motion, \Cref{Figure:Y} shows lateral, roll, and yaw responses, and for vertical platform motion \Cref{Figure:Z} shows vertical, fore-aft, and pitch responses. Appendix D shows gains, phases, and coherences for all 6 translational and rotational degrees of freedom for the pelvis, trunk, and head for all individual participants. The main effects are largely consistent between subjects with coherence generally exceeding 0.5 from 0.34-12 Hz. At the lowest frequency analyzed (0.17 Hz) coherence is low and hence these results are not very informative, presumably due to voluntary motion, non-linearity and limited perturbation power. From 0.34 Hz to about 2 Hz we see a gain close to 1 for the fore-aft direction with some amplification for the head. Gains are close to 1 from 0.34-3 Hz for the vertical direction. For the lateral direction, the response is not as straightforward as for the other directions and shows amplification around 1 Hz for the head and trunk with an additional peak around 3 Hz for the trunk.  
For all motion directions the phase for main responses at low frequency is close to zero indicating a limited timing difference between seat and body motion. Hence, at low frequencies the pelvis, trunk, and head translational motions closely follow the seat motion. At higher frequencies we see oscillations evidenced by gain peaks which are prominent in particular for vertical.

\par The main interactions shown in \Cref{Figure:X,Figure:Y,Figure:Z} are consistent between participants and show good coherence. Other interactions can be found in Appendix D and show partially inconsistent responses with a low coherence and variable phase which was largely expected. These interactions include lateral and roll responses to fore-aft and vertical seat motion which should be zero if the human body would be symmetric and would be perfectly aligned with the seat. Hence these non-zero interactions presumably represent postural and/or biomechanical asymmetries.

\subsection{Back support height and sitting posture} 
\setcounter{totalnumber}{4}

\begin{sidewaystable} 
 \renewcommand{\thetable}{\arabic{table}a}
\caption{Body segment peak response gains and frequencies for different heights of back support (Low, Middle and High), sitting postures (Erect, Pref:Preferred and Slouched) and 3 extra conditions of Eyes Closed (EC), Head Down (HD) and Low Amplitude (LA). Average and standard deviation for all subjects. Related statistics are in Table 1.b. \label{Table:TRdataP}}  
\scriptsize 
\label{Table:TRdata} 
\setlength\tabcolsep{2pt} 
\renewcommand{\arraystretch}{1.2} 
\begin{tabular}{@{}rlllclllclllclll@{}}\toprule 
& \multicolumn{3}{c}{$Low back support$} & \phantom{abc}& \multicolumn{3}{c}{$Middle back support$} & \phantom{abc}& 
\multicolumn{3}{c}{$High back support$} & \phantom{abc}& \multicolumn{3}{c}{$Extra conditions$}\\ 
\cmidrule{2-4} \cmidrule{6-8} \cmidrule{10-12} \cmidrule{14-16} 
& $Erect$ & $Pref.$ & $Slouched$ && $Erect$ & $Pref.$ & $Slouched$ && $Erect$ & $Pref.$ & $Slouched$ && $EC$ & $HD$ & $LA$  \\ \midrule
\multicolumn{2}{c}{\textbf{Fore-aft Perturbation - Fore-aft Response }}\\ 
Head Gain & 2.02 ± 0.4  & 1.98 ± 0.4  & 1.80 ± 0.4  && 1.88 ± 0.4  & 1.75 ± 0.3  & 1.71 ± 0.3  && 1.88 ± 0.4  & 1.74 ± 0.5  & 1.79 ± 0.3  && 3.15 ± 1.2  & 2.50 ± 0.6  & 3.52 ± 1.5 \footnotemark[2] \\ 
Freq. & 1.12 ± 0.2  & 1.32 ± 0.3  & 1.34 ± 0.3  && 1.59 ± 0.8  & 1.98 ± 1.1  & 2.14 ± 1.3  && 2.34 ± 1.2  & 2.50 ± 1.2  & 3.07 ± 1.5  && 3.78 ± 0.5  & 2.44 ± 1.0  & 4.42 ± 0.5 \footnotemark[2] \\ 
Trunk Gain & 1.38 ± 0.2  & 1.40 ± 0.2 \footnotemark[1]  & 1.31 ± 0.2  && 2.14 ± 0.5  & 2.48 ± 0.8  & 2.97 ± 0.6  && 2.70 ± 0.9  & 3.11 ± 0.7  & 3.45 ± 0.5  && 2.46 ± 0.7  & 2.17 ± 0.6  & 3.04 ± 0.9 \footnotemark[2] \\ 
Freq. & 0.97 ± 0.3  & 1.17 ± 0.3 \footnotemark[1]  & 1.26 ± 0.3  && 3.10 ± 0.9  & 3.44 ± 0.7  & 3.90 ± 0.3  && 3.33 ± 0.9  & 3.90 ± 0.3  & 4.09 ± 0.4  && 3.57 ± 0.5  & 3.30 ± 1.0  & 4.15 ± 0.5 \footnotemark[2] \\ 
Pelvis Gain & 1.52 ± 0.4  & 1.47 ± 0.4  & 1.41 ± 0.3 \footnotemark[1]  && 1.64 ± 0.4  & 1.61 ± 0.5  & 1.78 ± 0.5  && 1.69 ± 0.4  & 1.79 ± 0.4  & 1.86 ± 0.5  && 1.76 ± 0.5 \footnotemark[1]  & 1.79 ± 0.4 \footnotemark[1]  & 1.81 ± 0.4 \footnotemark[2] \\ 
Freq. & 5.30 ± 1.2  & 4.35 ± 1.0  & 3.59 ± 0.7 \footnotemark[1]  && 5.27 ± 0.9  & 4.39 ± 1.6  & 4.09 ± 1.2  && 4.94 ± 0.8  & 3.97 ± 0.5  & 3.87 ± 0.4  && 3.84 ± 0.3 \footnotemark[1]  & 3.92 ± 0.5 \footnotemark[1]  & 4.39 ± 0.3 \footnotemark[2] \\ 
\multicolumn{2}{l}{\textbf{Fore-aft Perturbation - Pitch Response}}\\ 
Head Gain & 17.45 ± 3.5  & 22.13 ± 6.4  & 25.67 ± 9.7  && 25.21 ± 6.9  & 25.40 ± 10.3  & 27.10 ± 6.6  && 28.61 ± 7.4  & 29.88 ± 6.6  & 30.39 ± 5.6  && 25.35 ± 10.1  & 24.41 ± 4.9  & 34.27 ± 10.4 \footnotemark[2] \\ 
Freq. & 4.06 ± 0.8  & 4.00 ± 0.6  & 4.33 ± 1.3  && 3.93 ± 0.7  & 4.14 ± 0.9  & 3.94 ± 0.4  && 3.81 ± 0.4  & 3.93 ± 0.3  & 4.30 ± 0.5  && 4.17 ± 1.1  & 3.81 ± 0.3  & 4.32 ± 0.5 \footnotemark[2] \\ 
Trunk Gain & 13.01 ± 3.9  & 13.20 ± 6.2  & 12.62 ± 5.9  && 12.31 ± 2.9  & 11.81 ± 6.1  & 10.92 ± 8.1  && 10.10 ± 4.2  & 8.99 ± 5.4  & 9.97 ± 8.4  && 10.47 ± 4.2  & 9.79 ± 4.1  & 13.31 ± 7.1 \footnotemark[2] \\ 
Freq. & 5.91 ± 0.7  & 5.40 ± 0.9  & 4.84 ± 1.0  && 5.37 ± 0.9  & 4.67 ± 1.0  & 5.06 ± 1.2  && 4.82 ± 0.8  & 5.24 ± 1.3  & 5.39 ± 1.3  && 4.60 ± 0.9  & 4.82 ± 0.9  & 5.10 ± 1.1 \footnotemark[2] \\ 
Pelvis Gain & 9.62 ± 6.2 \footnotemark[1]  & 9.64 ± 6.1  & 8.32 ± 6.0  && 13.19 ± 8.9  & 9.27 ± 5.9  & 11.71 ± 8.9  && 10.86 ± 8.4  & 10.59 ± 9.3  & 10.91 ± 8.3  && 8.68 ± 6.3 \footnotemark[1]  & 9.15 ± 8.1  & 10.64 ± 8.3 \footnotemark[2] \\ 
Freq. & 6.21 ± 1.0 \footnotemark[1]  & 5.76 ± 0.9  & 5.37 ± 1.1  && 5.94 ± 1.0  & 5.19 ± 0.9  & 4.94 ± 1.2  && 5.86 ± 1.2  & 5.00 ± 1.1  & 5.39 ± 1.5  && 5.17 ± 1.0 \footnotemark[1]  & 5.10 ± 1.2  & 5.75 ± 1.1 \footnotemark[2] \\ 
\hline 
\multicolumn{2}{l}{\textbf{Lateral Perturbation - Lateral Response}}\\ 
Head Gain & 1.85 ± 0.4  & 1.86 ± 0.3  & 2.04 ± 0.5  && 1.97 ± 0.4  & 1.98 ± 0.4 \footnotemark[1]  & 2.06 ± 0.6  && 1.99 ± 0.4  & 1.90 ± 0.4 \footnotemark[1]  & 2.01 ± 0.5  && 1.83 ± 0.4  & 1.94 ± 0.4  & 2.04 ± 0.7 \footnotemark[2] \\ 
Freq. & 0.82 ± 0.2  & 0.80 ± 0.2  & 1.03 ± 0.3  && 1.01 ± 0.2  & 1.02 ± 0.3 \footnotemark[1]  & 1.15 ± 0.2  && 1.07 ± 0.2  & 1.11 ± 0.2 \footnotemark[1]  & 1.21 ± 0.3  && 1.55 ± 0.8  & 1.32 ± 0.7  & 1.79 ± 0.8 \footnotemark[2] \\ 
Trunk Gain & 1.41 ± 0.2  & 1.49 ± 0.2  & 1.74 ± 0.3  && 1.61 ± 0.4  & 1.81 ± 0.3  & 1.97 ± 0.3  && 1.73 ± 0.5  & 1.92 ± 0.4  & 2.06 ± 0.3  && 2.17 ± 0.6  & 2.17 ± 0.5  & 2.87 ± 0.6 \footnotemark[2] \\ 
Freq. & 1.29 ± 0.9  & 1.99 ± 1.0  & 2.46 ± 1.1  && 2.04 ± 0.8  & 2.46 ± 0.6  & 2.35 ± 0.6  && 2.08 ± 0.9  & 2.32 ± 0.6  & 2.43 ± 0.5  && 2.69 ± 0.5  & 2.69 ± 0.5  & 2.98 ± 0.4 \footnotemark[2] \\ 
Pelvis Gain & 1.16 ± 0.1  & 1.17 ± 0.1  & 1.24 ± 0.2  && 1.17 ± 0.1  & 1.19 ± 0.1  & 1.27 ± 0.1  && 1.18 ± 0.1  & 1.16 ± 0.1  & 1.26 ± 0.1  && 1.18 ± 0.1  & 1.22 ± 0.1  & 1.24 ± 0.1 \footnotemark[2] \\ 
Freq. & 2.54 ± 0.6  & 2.31 ± 0.3  & 2.38 ± 0.5  && 2.44 ± 0.4  & 2.44 ± 0.5  & 2.22 ± 0.4  && 2.49 ± 0.4  & 2.29 ± 0.4  & 2.17 ± 0.5  && 2.35 ± 0.3  & 2.37 ± 0.3  & 2.65 ± 0.3 \footnotemark[2] \\ 
\multicolumn{2}{l}{\textbf{Lateral Perturbation - Roll Response}}\\ 
Head Gain & 13.53 ± 4.2  & 13.79 ± 5.0  & 14.64 ± 3.9  && 14.10 ± 3.9  & 14.93 ± 3.5  & 15.20 ± 3.9  && 14.79 ± 5.5  & 16.20 ± 4.8  & 15.40 ± 3.1  && 16.66 ± 4.5  & 14.28 ± 3.5  & 20.58 ± 7.3 \footnotemark[2] \\ 
Freq. & 2.83 ± 0.6  & 2.72 ± 0.4  & 2.59 ± 0.5  && 2.60 ± 0.4  & 2.43 ± 0.3  & 2.34 ± 0.4  && 2.54 ± 0.4  & 2.43 ± 0.2  & 2.29 ± 0.4  && 2.56 ± 0.4  & 2.62 ± 0.4  & 2.79 ± 0.3 \footnotemark[2] \\ 
Trunk Gain & 8.82 ± 1.5  & 9.02 ± 2.2  & 8.93 ± 2.1  && 9.36 ± 2.0  & 9.35 ± 1.8  & 8.34 ± 1.5  && 9.34 ± 2.3  & 9.17 ± 2.2  & 8.90 ± 2.3  && 9.13 ± 1.8  & 9.85 ± 1.8  & 10.35 ± 2.3 \footnotemark[2] \\ 
Freq. & 2.66 ± 0.3  & 2.86 ± 0.4  & 2.93 ± 0.9  && 2.78 ± 0.6  & 3.11 ± 0.8  & 2.77 ± 0.7  && 2.99 ± 0.5  & 3.11 ± 0.9  & 2.96 ± 0.5  && 2.75 ± 0.7  & 2.93 ± 0.6  & 3.11 ± 0.6 \footnotemark[2] \\ 
Pelvis Gain & 3.25 ± 2.0  & 2.90 ± 1.3  & 3.33 ± 1.1  && 2.56 ± 1.0  & 2.61 ± 1.0  & 3.44 ± 1.1  && 2.66 ± 1.3  & 3.03 ± 1.1  & 3.74 ± 1.2  && 2.98 ± 1.2  & 3.10 ± 1.5  & 3.64 ± 1.7 \footnotemark[2] \\ 
Freq. & 3.47 ± 1.1  & 3.29 ± 1.0  & 3.47 ± 1.0  && 3.50 ± 0.8  & 3.45 ± 0.9  & 3.41 ± 0.9  && 3.39 ± 0.9  & 3.20 ± 1.0  & 3.66 ± 1.1  && 3.10 ± 0.7  & 3.14 ± 0.7  & 3.52 ± 0.7 \footnotemark[2] \\ 
\hline 
\multicolumn{2}{l}{\textbf{Vertical Perturbation - Vertical Response}}\\ 
Head Gain & 4.30 ± 1.1  & 4.09 ± 1.0  & 3.54 ± 0.8  && 4.50 ± 0.6  & 4.40 ± 1.0  & 4.48 ± 0.5  && 4.30 ± 0.5  & 4.46 ± 0.6  & 4.42 ± 0.6  && 4.92 ± 0.7  & 4.76 ± 0.7  & 5.04 ± 0.7 \footnotemark[2] \\ 
Freq. & 4.78 ± 0.3  & 4.90 ± 0.5  & 5.19 ± 0.5  && 4.60 ± 0.2  & 4.49 ± 0.3  & 4.58 ± 0.2  && 4.64 ± 0.3  & 4.49 ± 0.3  & 4.37 ± 0.2  && 4.54 ± 0.2  & 4.94 ± 0.5  & 4.83 ± 0.3 \footnotemark[2] \\ 
Trunk Gain & 4.43 ± 1.3  & 4.27 ± 1.2  & 4.24 ± 1.2  && 4.44 ± 1.0  & 4.30 ± 1.1  & 4.98 ± 0.9  && 4.27 ± 0.9  & 4.66 ± 0.9  & 5.09 ± 0.9  && 4.47 ± 0.9  & 4.07 ± 0.8  & 4.75 ± 0.7 \footnotemark[2] \\ 
Freq. & 4.99 ± 0.7  & 4.75 ± 0.7  & 5.16 ± 0.5  && 4.57 ± 0.2  & 4.49 ± 0.2  & 4.57 ± 0.3  && 4.60 ± 0.3  & 4.43 ± 0.3  & 4.39 ± 0.2  && 4.43 ± 0.2  & 4.42 ± 0.5  & 4.73 ± 0.3 \footnotemark[2] \\ 
Pelvis Gain & 2.00 ± 0.9  & 2.27 ± 0.6  & 2.53 ± 0.7  && 1.99 ± 0.6  & 2.09 ± 0.6  & 2.16 ± 0.5  && 2.05 ± 0.5  & 2.23 ± 0.5  & 2.29 ± 0.8  && 2.39 ± 0.6  & 2.44 ± 0.6  & 2.61 ± 0.7 \footnotemark[2] \\ 
Freq. & 4.72 ± 1.0  & 4.99 ± 1.2  & 4.94 ± 0.9  && 4.61 ± 0.6  & 4.33 ± 0.8  & 4.66 ± 0.5  && 4.46 ± 0.4  & 4.36 ± 0.4  & 4.37 ± 0.3  && 4.32 ± 1.0  & 5.09 ± 0.8  & 5.10 ± 0.8 \footnotemark[2] \\ 
\multicolumn{2}{l}{\textbf{Vertical Perturbation - Pitch Response}}\\ 
Head Gain & 23.85 ± 9.1  & 22.03 ± 8.2  & 16.47 ± 6.4  && 24.92 ± 9.9  & 23.05 ± 7.7  & 18.57 ± 10.7  && 26.18 ± 10.5  & 25.25 ± 9.8  & 20.44 ± 5.1  && 24.21 ± 10.3  & 35.33 ± 9.0  & 25.79 ± 6.9 \footnotemark[2] \\ 
Freq. & 4.93 ± 0.4  & 5.27 ± 0.8  & 5.77 ± 1.0  && 4.96 ± 0.8  & 4.94 ± 0.7  & 5.70 ± 0.8  && 5.07 ± 0.7  & 5.10 ± 0.7  & 5.67 ± 0.9  && 5.16 ± 0.7  & 4.87 ± 0.4  & 5.36 ± 0.5 \footnotemark[2] \\ 
Trunk Gain & 15.19 ± 8.2  & 18.24 ± 12.2  & 18.62 ± 12.2  && 17.54 ± 10.1  & 16.20 ± 8.1 \footnotemark[1]  & 17.15 ± 9.2  && 19.42 ± 10.5  & 16.48 ± 5.6  & 17.84 ± 9.3  && 16.92 ± 9.2  & 19.12 ± 16.6  & 21.68 ± 13.0 \footnotemark[2] \\ 
Freq. & 6.19 ± 0.9  & 6.68 ± 0.9  & 6.22 ± 1.1  && 6.40 ± 1.1  & 5.97 ± 1.1 \footnotemark[1]  & 5.94 ± 1.0  && 6.46 ± 0.9  & 6.52 ± 1.2  & 6.86 ± 0.9  && 5.94 ± 1.2  & 6.18 ± 0.8  & 6.31 ± 0.7 \footnotemark[2] \\ 
Pelvis Gain & 6.57 ± 4.2 \footnotemark[3]  & 8.19 ± 4.7  & 7.26 ± 5.2 \footnotemark[1]  && 7.75 ± 4.8 \footnotemark[1]  & 7.65 ± 5.0  & 14.60 ± 8.7  && 6.22 ± 3.8  & 7.77 ± 7.0  & 11.53 ± 7.2  && 6.80 ± 4.7  & 7.22 ± 5.6  & 7.38 ± 5.5 \footnotemark[2] \\ 
Freq. & 5.31 ± 0.8 \footnotemark[3]  & 6.03 ± 1.1  & 5.86 ± 0.8 \footnotemark[1]  && 5.95 ± 1.3 \footnotemark[1]  & 5.49 ± 1.4  & 5.04 ± 0.5  && 5.45 ± 1.0  & 5.73 ± 1.1  & 5.92 ± 1.4  && 5.63 ± 1.1  & 5.55 ± 0.7  & 6.07 ± 1.2 \footnotemark[2] \\ 
\hline 
\bottomrule 
\end{tabular} 
\footnotetext[1]{Peaks not found for 1 subjects} 
\footnotetext[2]{Peaks not found for 2 subjects} 
\footnotetext[3]{Peaks not found for 3 subjects} 
\end{sidewaystable}
\begin{table}
  \addtocounter{table}{-1}
 \renewcommand{\thetable}{\arabic{table}.b}
\caption{Significance of results in Table 1a. P values of post hoc tests for main translational and rotational responses. E:Erect, P:Preferred, S:Slouched, L:Low, M:Middle, H:High.  \label{Table:StatsP}} 
\label{Table:StatsP}
\begin{tabular}{lllllclllclll} 
\toprule
\begin{tabular}[c]{@{}l@{}}\\\\\end{tabular} &       & \multicolumn{3}{c}{Posture} & \multicolumn{1}{l}{} & \multicolumn{3}{c}{Support} &                      & \multicolumn{3}{c}{Extra}             \\ 
\midrule
                                             &       & E-P   & E-S   & S-P         &                      & L-H   & M-H   & L-M         &                      & EC-MP & HD-MP & LA-MP                 \\ 
\midrule
\multicolumn{2}{l}{Fore-aft Perturbation}            & \multicolumn{7}{l}{Fore-aft Response}                                            & \multicolumn{1}{l}{} &       &       & \multicolumn{1}{c}{}  \\
Head                                         & Gain  & 0.700 & 0.165 & 1           &                      & 1     & 0.747 & 0.511       &                      & 0.008 & 0.008 & 0.039                 \\
                                             & Freq. & 0.509 & 1     & 0.509       &                      & 0.049 & 0.133 & 0.069       &                      & 0.001 & 0.789 & 0.001                 \\
Trunk                                        & Gain  & 0.062 & 0.020 & 0.3932      &                      & 0.002 & 0.125 & 0.008       &                      & 1     & 0.675 & 0.084                 \\
                                             & Freq. & 0.160 & 0.088 & 0.268       &                      & 0.001 & 1     & 0.001       &                      & 1     & 1     & 0.035                 \\
Pelvis                                       & Gain  & 1     & 1     & 1           &                      & 0.018 & 0.463 & 0.275       &                      & 0.923 & 0.530 & 0.251                 \\
                                             & Freq. & 0.001 & 0.001 & 0.528       &                      & 0.635 & 1     & 0.175       &                      & 1     & 1     & 1                     \\
\multicolumn{2}{l}{}                                 & \multicolumn{7}{l}{Pitch Response}                                               & \multicolumn{1}{l}{} &       &       & \multicolumn{1}{c}{}  \\
Head                                         & Gain  & 1     & 1     & 1           &                      & 0.018 & 0.181 & 0.013       &                      & 1     & 1     & 0.638                 \\
                                             & Freq. & 1     & 1     & 1           &                      & 1     & 1     & 1           &                      & 1     & 1     & 1                     \\
Trunk                                        & Gain  & 1     & 1     & 1           &                      & 0.108 & 0.087 & 1           &                      & 1     & 0.640 & 1                     \\
                                             & Freq. & 1     & 1     & 1           &                      & 0.452 & 1     & 1           &                      & 1     & 1     & 1                     \\
Pelvis                                       & Gain  & 0.374 & 1     & 0.471       &                      & 1     & 1     & 1           &                      & 1     & 1     & 1                     \\
                                             & Freq. & 0.509 & 0.092 & 0.609       &                      & 0.845 & 0.791 & 1           &                      & 1     & 1     & 1                     \\ 
\hline
\multicolumn{2}{l}{Lateral Perturbation}             & \multicolumn{7}{l}{Lateral Response}                                             & \multicolumn{1}{l}{} &       &       & \multicolumn{1}{c}{}  \\
Head                                         & Gain  & 1     & 0.644 & 0.864       &                      & 0.485 & 1     & 0.418       &                      & 1     & 1     & 1                     \\
                                             & Freq. & 1     & 0.340 & 0.023       &                      & 0.006 & 0.142 & 0.004       &                      & 0.063 & 0.767 & 0.004                 \\
Trunk                                        & Gain  & 0.164 & 0.180 & 0.602       &                      & 0.034 & 0.698 & 0.556       &                      & 0.007 & 0.236 & 0.001                 \\
                                             & Freq. & 0.243 & 0.271 & 1           &                      & 0.347 & 0.397 & 0.936       &                      & 1     & 0.894 & 0.055                 \\
Pelvis                                       & Gain  & 1     & 1     & 0.973       &                      & 1     & 1     & 1           &                      & 1     & 1     & 0.242                 \\
                                             & Freq. & 1     & 0.593 & 0.566       &                      & 0.181 & 0.775 & 0.285       &                      & 1     & 1     & 0.551                 \\
\multicolumn{2}{l}{}                                 & \multicolumn{7}{l}{Roll Response}                                                & \multicolumn{1}{l}{} &       &       & \multicolumn{1}{c}{}  \\
Head                                         & Gain  & 1     & 1     & 1           &                      & 0.406 & 1     & 0.715       &                      & 0.444 & 0.569 & 1                     \\
                                             & Freq. & 0.332 & 0.015 & 0.261       &                      & 0.005 & 0.857 & 0.008       &                      & 1     & 0.994 & 0.003                 \\
Trunk                                        & Gain  & 1     & 0.924 & 1           &                      & 1     & 1     & 0.489       &                      & 1     & 1     & 0.172                 \\
                                             & Freq. & 0.048 & 0.280 & 1           &                      & 1     & 1     & 1           &                      & 1     & 0.898 & 1                     \\
Pelvis                                       & Gain  & 1     & 0.439 & 0.358       &                      & 0.511 & 1     & 0.208       &                      & 1     & 1     & 1                     \\
                                             & Freq. & 0.686 & 1     & 0.461       &                      & 1     & 1     & 1           &                      & 1     & 1     & 0.135                 \\ 
\hline
\multicolumn{2}{l}{Vertical Perturbation}            & \multicolumn{7}{l}{Vertical Response}                                            & \multicolumn{1}{l}{} &       &       & \multicolumn{1}{c}{}  \\
Head                                         & Gain  & 1     & 1     & 1           &                      & 1     & 0.269 & 1           &                      & 0.579 & 1     & 0.215                 \\
                                             & Freq. & 1     & 1     & 1           &                      & 0.045 & 1     & 0.118       &                      & 1     & 0.113 & 0.003                 \\
Trunk                                        & Gain  & 0.103 & 0.106 & 0.216       &                      & 1     & 0.824 & 1           &                      & 1     & 1     & 0.598                 \\
                                             & Freq. & 1     & 1     & 1           &                      & 0.017 & 0.121 & 0.077       &                      & 1     & 1     & 0.006                 \\
Pelvis                                       & Gain  & 0.113 & 0.039 & 0.162       &                      & 1     & 1     & 1           &                      & 1     & 1     & 0.536                 \\
                                             & Freq. & 1     & 1     & 1           &                      & 0.104 & 0.051 & 0.244       &                      & 1     & 0.189 & 0.235                 \\
\multicolumn{2}{l}{}                                 & \multicolumn{7}{l}{Pitch Response}                                               & \multicolumn{1}{l}{} &       &       & \multicolumn{1}{c}{}  \\
Head                                         & Gain  & 0.193 & 0.012 & 0.099       &                      & 0.663 & 0.768 & 1           &                      & 1     & 0.005 & 1                     \\
                                             & Freq. & 0.503 & 0.001 & 0.001       &                      & 1     & 1     & 1           &                      & 0.295 & 1     & 0.025                 \\
Trunk                                        & Gain  & 1     & 1     & 1           &                      & 1     & 0.453 & 1           &                      & 1     & 1     & 0.456                 \\
                                             & Freq. & 1     & 1     & 1           &                      & 1     & 0.083 & 0.027       &                      & 1     & 0.326 & 1                     \\
Pelvis                                       & Gain  & 0.324 & 0.245 & 0.024       &                      & 1     & 1     & 1           &                      & 1     & 1     & 1                     \\
                                             & Freq. & 0.368 & 1     & 0.212       &                      & 0.408 & 1     & 0.752       &                      & 1     & 1     & 1                     \\
\bottomrule
\end{tabular}
\end{table}
\begin{table} 
 \renewcommand{\thetable}{\arabic{table}a}
\caption{Average gain response between 1 and 2 Hz for different heights of back support (Low, Middle and High), sitting postures (Erect, Pref:Preferred and Slouched) and 3 extra conditions of Eyes Closed (EC), Head Down (HD) and Low Amplitude (LA). Average and standard deviation for all subjects. Related statistics are in Table 2.b. \label{Table:TRdataAvg}}  
\scriptsize 
\label{Table:TRdata} 
\setlength\tabcolsep{1.7pt} 
\renewcommand{\arraystretch}{1.2} 
\hspace*{-0.1cm} 
\begin{tabular}{@{}llrrclcrclcrclll@{}}\toprule 
& \multicolumn{3}{c}{$Low$} & \phantom{abc}& \multicolumn{3}{c}{$Middle$} & \phantom{abc}& 
\multicolumn{3}{c}{$High$} & \phantom{abc}& \multicolumn{3}{c}{$Extra$}\\ 
\cmidrule{2-4} \cmidrule{6-8} \cmidrule{10-12} \cmidrule{14-16} 
& $Erect$ & $Pref.$ & $Slouched$ && $Erect$ & $Pref.$ & $Slouched$ && $Erect$ & $Pref.$ & $Slouched$ && $EC$ & $HD$ & $LA$  \\ \midrule
\noalign{\vskip 1.1mm} 
\multicolumn{4}{l}{\textbf{Fore-aft Perturbation - Fore-aft Response }}\\ 
\noalign{\vskip 1.1mm} 
Head & 1.6±0.3  & 1.7±0.3  & 1.5±0.3  && 1.5±0.3  & 1.4±0.3  & 1.3±0.3  && 1.4±0.4  & 1.3±0.3  & 1.2±0.2  && 1.3±0.2  & 1.7±0.3  & 1.1±0.4 \\ 
Trunk & 1.0±0.2  & 1.2±0.2  & 1.2±0.1  && 1.3±0.1  & 1.2±0.1  & 1.2±0.1  && 1.2±0.1  & 1.2±0.1  & 1.2±0.1  && 1.2±0.1  & 1.4±0.2  & 1.1±0.1 \\ 
Pelvis & 0.9±0.1  & 1.0±0.1  & 1.0±0.1  && 0.9±0.1  & 1.0±0.1  & 1.0±0.1  && 0.9±0.1  & 1.0±0.1  & 1.0±0.1  && 1.0±0.1  & 1.0±0.1  & 1.0±0.1 \\ 
\noalign{\vskip 1.1mm} 
\multicolumn{4}{l}{\textbf{Fore-aft Perturbation - Pitch Response}}\\ 
\noalign{\vskip 1.1mm} 
Head & 7.5±2.2  & 8.4±2.0  & 9.2±2.9  && 9.3±3.0  & 8.6±2.4  & 8.4±2.4  && 8.3±2.1  & 8.1±2.0  & 8.4±2.0  && 10.2±2.8  & 6.8±2.1  & 8.3±3.5 \\ 
Trunk & 3.4±0.7  & 2.9±1.4  & 1.7±0.8  && 1.7±1.0  & 1.3±0.5  & 0.8±0.3  && 1.3±0.9  & 0.9±0.5  & 0.6±0.3  && 1.2±0.7  & 1.8±0.7  & 1.3±0.5 \\ 
Pelvis  & 1.0±1.4  & 0.6±0.3  & 0.4±0.4  && 0.4±0.2  & 0.4±0.3  & 0.6±0.5  && 0.4±0.3  & 0.4±0.4  & 0.5±0.4  && 0.4±0.2  & 0.5±0.2  & 0.4±0.2 \\ 
\hline 
\noalign{\vskip 1.1mm} 
\multicolumn{4}{l}{\textbf{Lateral Perturbation - Lateral Response}}\\ 
\noalign{\vskip 1.1mm} 
Head & 1.3±0.3  & 1.3±0.2  & 1.5±0.4  && 1.4±0.3  & 1.5±0.3  & 1.5±0.4  && 1.5±0.3  & 1.5±0.3  & 1.5±0.3  && 1.4±0.4  & 1.4±0.3  & 1.4±0.4 \\ 
Trunk & 0.8±0.2  & 0.9±0.2  & 1.1±0.2  && 1.0±0.1  & 1.2±0.2  & 1.4±0.2  && 1.1±0.2  & 1.3±0.2  & 1.5±0.2  && 1.3±0.1  & 1.2±0.1  & 1.3±0.1 \\ 
Pelvis & 1.0±0.0  & 1.0±0.0  & 1.1±0.1  && 1.0±0.0  & 1.0±0.1  & 1.1±0.1  && 1.0±0.0  & 1.0±0.1  & 1.1±0.1  && 1.0±0.1  & 1.1±0.1  & 1.0±0.0 \\ 
\noalign{\vskip 1.1mm} 
\multicolumn{4}{l}{\textbf{Lateral Perturbation - Roll Response}}\\ 
\noalign{\vskip 1.1mm} 
Head & 6.5±2.2  & 7.3±2.2  & 8.2±2.0  && 7.6±2.0  & 8.4±2.0  & 9.6±2.7  && 8.2±2.2  & 9.4±2.3  & 9.7±2.0  && 10.5±3.1  & 7.8±2.0  & 9.8±3.6 \\ 
Trunk & 4.9±0.7  & 5.1±0.9  & 5.0±1.4  && 4.9±0.8  & 4.5±1.3  & 4.0±1.0  && 4.7±1.2  & 4.3±1.3  & 3.9±1.1  && 4.8±1.1  & 5.0±1.2  & 4.5±1.2 \\ 
Pelvis & 0.7±0.7  & 0.7±0.4  & 0.9±0.5  && 0.5±0.2  & 0.6±0.3  & 1.0±0.4  && 0.5±0.2  & 0.8±0.3  & 1.1±0.5  && 0.7±0.3  & 0.7±0.3  & 0.7±0.3 \\ 
\hline 
\noalign{\vskip 1.1mm} 
\multicolumn{4}{l}{\textbf{Vertical Perturbation - Vertical Response}}\\ 
\noalign{\vskip 1.1mm} 
Head & 1.1±0.0  & 1.1±0.0  & 1.1±0.0  && 1.1±0.0  & 1.1±0.0  & 1.1±0.0  && 1.1±0.0  & 1.1±0.0  & 1.1±0.0  && 1.1±0.0  & 1.1±0.0  & 1.1±0.0 \\ 
Trunk & 1.1±0.0  & 1.1±0.0  & 1.1±0.0  && 1.1±0.0  & 1.1±0.0  & 1.1±0.0  && 1.1±0.0  & 1.1±0.0  & 1.1±0.0  && 1.1±0.0  & 1.1±0.0  & 1.1±0.0 \\ 
Pelvis & 0.9±0.1  & 1.0±0.1  & 1.0±0.1  && 0.9±0.1  & 1.0±0.1  & 1.0±0.1  && 0.9±0.1  & 1.0±0.1  & 1.0±0.1  && 1.0±0.1  & 1.0±0.1  & 1.0±0.1 \\ 
\noalign{\vskip 1.1mm} 
\multicolumn{4}{l}{\textbf{Vertical Perturbation - Pitch Response}}\\ 
\noalign{\vskip 1.1mm} 
Head & 0.9±0.4  & 1.0±0.5  & 1.0±0.4  && 1.0±0.4  & 1.3±0.5  & 1.1±0.2  && 1.0±0.3  & 1.1±0.4  & 1.0±0.3  && 1.3±0.4  & 1.8±0.3  & 1.4±0.4 \\ 
Trunk & 0.3±0.1  & 0.3±0.2  & 0.3±0.1  && 0.3±0.1  & 0.4±0.2  & 0.3±0.1  && 0.3±0.1  & 0.3±0.1  & 0.3±0.1  && 0.3±0.2  & 0.4±0.2  & 0.5±0.2 \\ 
Pelvis  & 0.2±0.1  & 0.2±0.1  & 0.2±0.1  && 0.2±0.1  & 0.2±0.2  & 0.4±0.3  && 0.2±0.1  & 0.2±0.2  & 0.3±0.2  && 0.2±0.2  & 0.2±0.2  & 0.2±0.2 \\ 
\hline 
\bottomrule 
\end{tabular} 
\end{table}
\begin{table}[]
  \addtocounter{table}{-1}
 \renewcommand{\thetable}{\arabic{table}.b}
\caption{Significance of results in Table 2a. P values of post hoc tests of main translational and rotational responses for average of gain between 1 and 2 Hz. E:Erect, P:Preferred, S:Slouched, L:Low M:Middle, H:High and 3 extra conditions of Eyes Closed (EC), Head Down (HD) and Low Amplitude (LA).   \label{Table:StatsAvg}}
\label{Table:StatsAvg}
\begin{tabular}{llllclllclll} 
\toprule
\begin{tabular}[c]{@{}l@{}}\\\\\end{tabular} & \multicolumn{3}{c}{Posture} & \multicolumn{1}{l}{} & \multicolumn{3}{c}{Support} &                      & \multicolumn{3}{c}{Extra}             \\ 
\midrule
                                             & E-P   & E-S   & S-P         &                      & L-H   & M-H   & L-M         &                      & EC-MP & HD-MP                & LA-MP  \\ 
\midrule
Fore-aft Perturbation                        & \multicolumn{7}{l}{Fore-aft Response}                                            & \multicolumn{1}{l}{} &       & \multicolumn{1}{c}{} &        \\
Head                                         & 1     & 0.074 & 0.219       &                      & 0.002 & 0.156 & 0.002       &                      & 1     & 0.045                & 0.750  \\
Trunk                                        & 0.426 & 0.794 & 1           &                      & 0.159 & 0.439 & 0.022       &                      & 0.371 & 0.011                & 0.012  \\
Pelvis                                       & 0.012 & 0.001 & 0.003       &                      & 0.246 & 0.355 & 1           &                      & 0.562 & 0.233                & 1      \\
                                             & \multicolumn{7}{l}{Pitch Response}                                               & \multicolumn{1}{l}{} &       & \multicolumn{1}{c}{} &        \\
Head                                         & 1     & 0.864 & 0.842       &                      & 1     & 0.895 & 1           &                      & 0.013 & 0.060                & 1      \\
Trunk                                        & 0.001 & 0.001 & 0.001       &                      & 0.001 & 0.041 & 0.001       &                      & 0.477 & 0.058                & 1      \\
Pelvis                                       & 1     & 1     & 1           &                      & 0.175 & 1     & 0.943       &                      & 1     & 1                    & 1      \\ 
\hline
Lateral Perturbation                         & \multicolumn{7}{l}{Lateral Response}                                             & \multicolumn{1}{l}{} &       & \multicolumn{1}{c}{} &        \\
Head                                         & 1     & 0.126 & 0.096       &                      & 0.042 & 1     & 0.056       &                      & 1     & 1                    & 1      \\
Trunk                                        & 0.001 & 0.001 & 0.001       &                      & 0.001 & 0.031 & 0.001       &                      & 0.065 & 1                    & 0.211  \\
Pelvis                                       & 0.866 & 0.003 & 0.004       &                      & 0.534 & 1     & 0.384       &                      & 1     & 1                    & 0.399  \\
                                             & \multicolumn{7}{l}{Roll Response}                                                & \multicolumn{1}{l}{} &       & \multicolumn{1}{c}{} &        \\
Head                                         & 0.005 & 0.001 & 0.032       &                      & 0.001 & 0.137 & 0.005       &                      & 0.001 & 1                    & 1      \\
Trunk                                        & 0.156 & 0.013 & 0.432       &                      & 0.021 & 1     & 0.023       &                      & 1     & 0.809                & 1      \\
Pelvis                                       & 0.033 & 0.001 & 0.013       &                      & 1     & 0.663 & 1           &                      & 1     & 1                    & 1      \\ 
\hline
Vertical Perturbation                        & \multicolumn{7}{l}{Vertical Response}                                            & \multicolumn{1}{l}{} &       & \multicolumn{1}{c}{} &        \\
Head                                         & 1     & 1     & 1           &                      & 0.001 & 0.722 & 0.004       &                      & 1     & 1                    & 0.625  \\
Trunk                                        & 0.038 & 0.244 & 1           &                      & 0.323 & 1     & 1           &                      & 1     & 1                    & 1      \\
Pelvis                                       & 0.038 & 0.006 & 0.007       &                      & 0.818 & 0.027 & 1           &                      & 0.587 & 0.213                & 1      \\
                                             & \multicolumn{7}{l}{Pitch Response}                                               & \multicolumn{1}{l}{} &       & \multicolumn{1}{c}{} &        \\
Head                                         & 0.014 & 0.052 & 1           &                      & 1     & 0.106 & 0.139       &                      & 1     & 0.008                & 1      \\
Trunk                                        & 0.275 & 1     & 1           &                      & 0.831 & 1     & 0.621       &                      & 1     & 0.992                & 0.016  \\
Pelvis                                       & 1     & 0.130 & 0.055       &                      & 1     & 0.633 & 0.138       &                      & 1     & 1                    & 1      \\
\bottomrule
\end{tabular}
\end{table}

\begin{figure}[p]
 \centering
\includegraphics[trim = 100 30 100 20,width=1\textwidth]{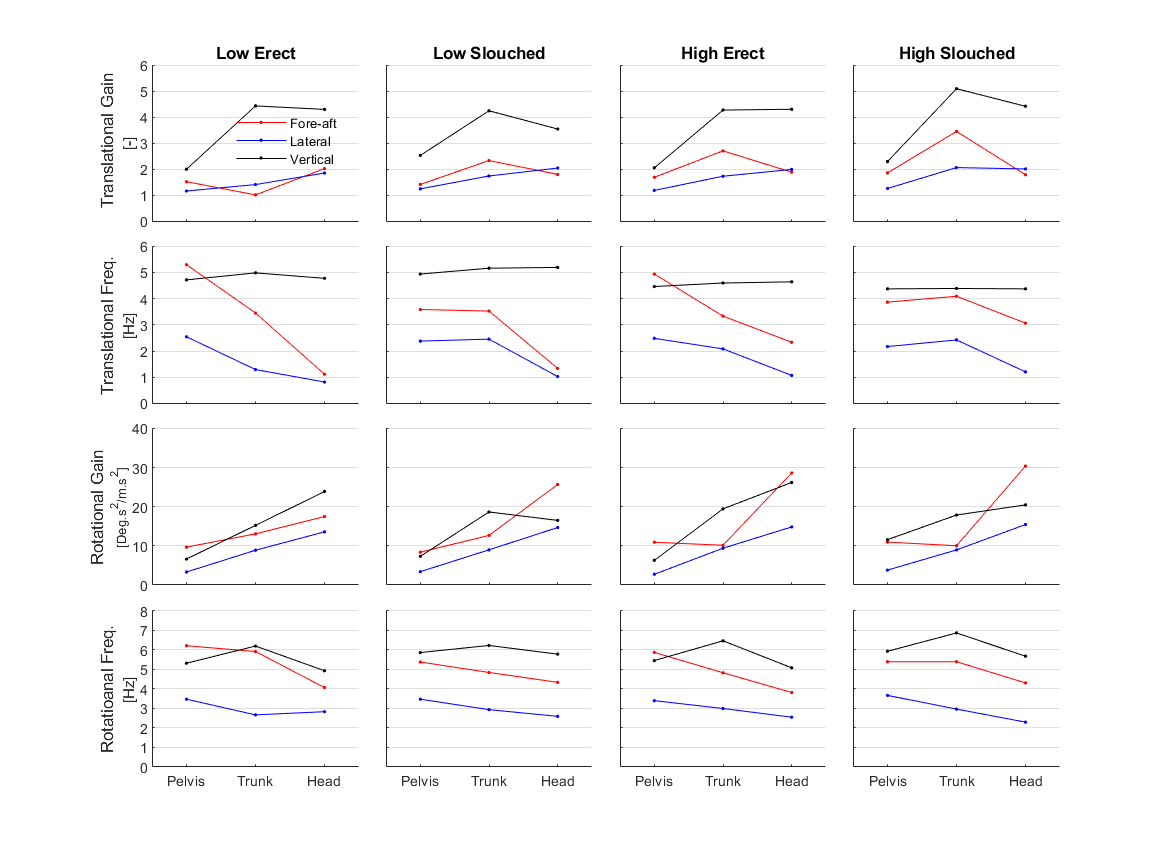}

\caption{Mean peak gains and related frequencies in transmission of vibrations from seat to head for selected sitting postures and back rest heights in main translational responses to perturbations (fore-aft to fore-aft, lateral to lateral, vertical to vertical) and main rotational responses (pitch to fore-aft, roll to lateral and pitch to vertical). For standard deviations, see \Cref{Table:TRdataP}. }
\label{Figure:PTH}
\end{figure}

\Cref{Table:TRdataP} presents the average and standard deviation of peaks and related frequencies of gain responses. Selected peak gains and frequencies as function of posture and support height are shown in \Cref{Figure:PTH}. The highest translational peak gains were found in vertical loading in trunk and head with peak gains between 3.6 and 5.1 between 4.5-5.2 Hz. The highest rotational gains were found in the head in all motion directions and in all conditions except for Low Slouched where the trunk rotation slightly exceeded the head rotation.   

\subsubsection{Peak translational responses}
\par  Analyzing all body segments jointly, peak gains of segment translational responses to platform vibration are influenced significantly by both back support height and sitting posture (support: F(2) = 15.771, p $<$ 0.001; posture: F(2) = 19.197, p $<$ 0.001) for the main translational response to each perturbation direction (fore-aft to fore-aft, lateral to lateral, and vertical to vertical). The peak gains are significantly lower for low support than for middle support (Table \ref{Table:TRdataP}, p = 0.014) and for high support (Table \ref{Table:TRdataP}, p = 0.001). There is no significant difference between middle and high support (p = 0.180). The slouched posture leads to higher peak gains than preferred (p = 0.014), and erect (p = 0.011). Between preferred and erect there is no significant difference (p = 0.360). Considering the significant effect of segment$\times$direction$\times$support (F(8) = 11.449, p \textless0.001) and segment$\times$direction$\times$posture (F(8) = 3.389, p \textless0.001), post hoc tests were performed to investigate effects on peak gains for each body segment in each direction. \Cref{Table:StatsP} summarizes the results for each direction and segment for both sitting posture and support. 
With fore-aft perturbations the trunk peak gain is significantly higher in slouched compared to erect, whereas trunk and pelvis peak gains are significantly higher with high and medium compared to low support.
With vertical perturbations the pelvis peak gain is significantly higher in slouched compared to erect whereas the head and trunk peak gains are significantly higher with high support.

\par 

    The peak frequencies are not modulated by back support height or sitting posture when all motion directions and segments are jointly analyzed (support: F(2) = 0.593, p = 0.560 ; posture: F(2) = 0.122, p = 0.884). The interactions of segment$\times$direction$\times$support (F(8) = 4.437, p \textless0.001) and segment$\times$direction$\times$posture (F(8) = 9.810, p \textless0.001) are significant which allows performing post hoc tests to investigate the effect of support and posture on peak frequencies for each segment in each  perturbation direction (Table \ref{Table:StatsP}). During fore-aft perturbations, peak frequencies are significantly affected by posture in the pelvis (\Cref{Table:StatsP}). 
    Pelvis peak frequencies are significantly higher for erect sitting in comparison with preferred and slouched, while there is no significant difference between preferred and slouched (\Cref{Table:TRdataP}). In the lateral direction, support height and posture have a significant effect on peak frequencies of lateral head acceleration. Peak frequencies are significantly higher in low support conditions in comparison with middle and high support (\Cref{Table:TRdataP,Table:StatsP}), while there is no significant difference between middle and high support. 
    During vertical vibration, only the peak frequencies for the head and trunk are influenced (\Cref{Table:StatsP}) where low support results in higher peak frequencies in comparison with high support.

\subsubsection{Peak rotational responses}
\par Main rotational peak gains (when all motion directions and segments are jointly analysed) are not modulated by either sitting posture (F(2) = 0.612, p = 0.657), or back support height (F(2) = 0.897, p = 0.476). Considering the significant effect of segment$\times$direction$\times$support (F(8) = 2.462, p = 0.010) and segment$\times$direction$\times$posture (F(8) = 4.358, p \textless 0.001), post hoc tests were performed to investigate effects for each segment in each direction (\Cref{Table:StatsP}). No significant effects of posture are found during fore-aft and lateral perturbations. However, during vertical perturbations, the head pitch response with erect sitting posture is significantly (around 40\%) higher than with slouched posture. High and middle back support lead to significantly higher head pitch gains than low support during fore-aft perturbations (\Cref{Table:TRdataP,Table:StatsP}).

    \par Similar to main rotational peak gains, peak frequencies are also not modulated by either sitting posture (F(2) = 0.136, p = 0.186), or backrest height (F(2) = 1.870, p = 0.186) when all motion directions and segments are jointly considered in the analysis. No significant interactions are found between direction, segment and support (F(8) = 1.0137, p = 0.434). However, interactions between direction, segment and support are significant (F(8) = 8.843, p $<$ 0.001).  During lateral perturbations, the head roll peak frequencies are significantly higher for erect sitting posture than slouched. During vertical, peak frequencies of head pitch were significantly lower for erect and preferred sitting posture than slouched (\Cref{Table:TRdataP,Table:StatsP}).

\subsubsection{Low frequency (1-2 Hz) translational and rotational responses}
\par Low frequency gains were analysed taking the average gain from 1-2 Hz where consistent and coherent responses are seen across participants while showing similar trends as even lower frequencies. \Cref{Table:TRdataAvg} provides 1-2 Hz gains for all motion directions and body segments and the related statistics are reported in  \Cref{Table:StatsAvg}. Effects of support and posture on translational responses are negligible. Effects of posture are significant for the pelvis during fore-aft and lateral perturbations, but the actual difference is rather low. Rotational responses, on the other hand, are modulated by support and posture particularly for trunk during fore-aft and for head during lateral excitation.


\subsection{Eyes closed}
\begin{figure}[p]
\centering
\includegraphics[width=0.85\textwidth]{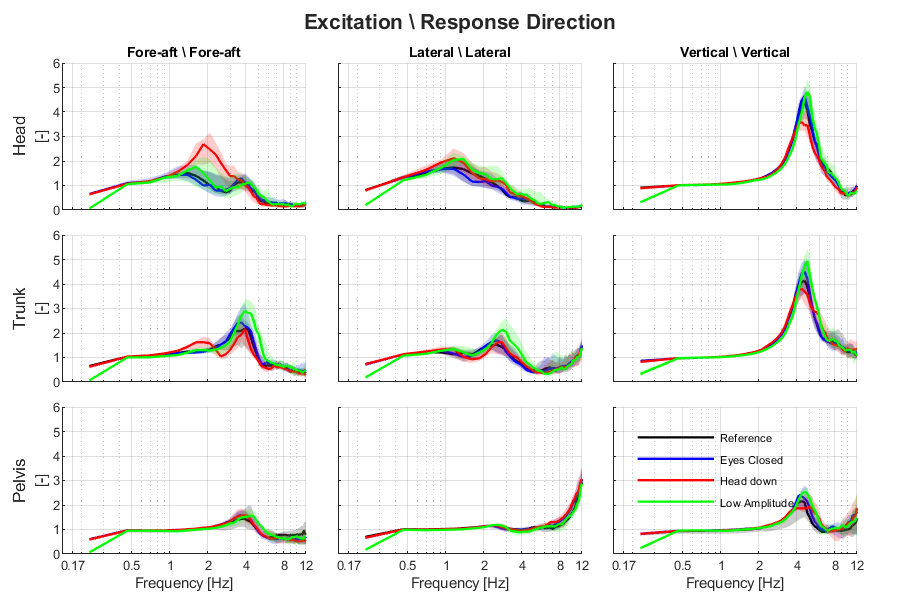}
\includegraphics[width=0.85\textwidth]{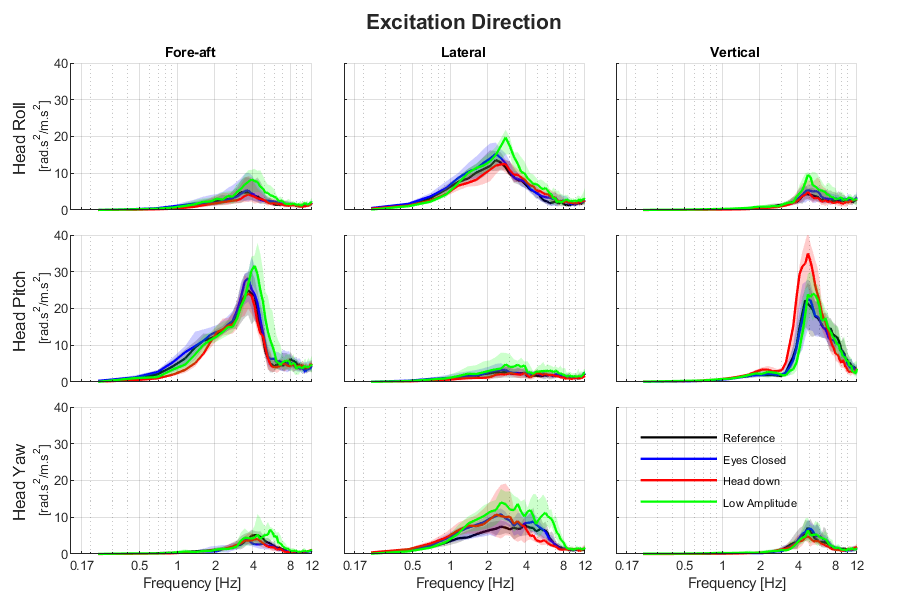}
\caption{Extra conditions for middle back support in preferred posture. Reference condition with vision, head looking forward and amplitude scale 1 (black line) without vision (blue line), Head Down (red line) and Low Amplitude (green line). 
Median gains (solid lines) with 25\textsuperscript{th} and 75\textsuperscript{th} percentile (shadows). 
Upper panel: main responses for head, trunk and pelvis (fore-aft response to fore-aft perturbation, lateral response to lateral perturbation, vertical response to vertical perturbation). 
Lower panel: Rotational responses for the head. 
Left column: fore-aft, mid: lateral, right: vertical perturbations.}
\label{Figure:extra}
\end{figure}

Figure \ref{Figure:extra} shows body segment responses with and without vision. 
Translational peak gains are significantly affected by vision  (F(1)= 11.799, p = 0.004). Considering the significant effect of segment$\times$direction$\times$condition (F(4)=9.823, p $<$ 0.001) post hoc tests were performed to investigate effects for each body segment in each direction. Post hoc tests show that the effect of vision on peak gain is only significant for head translation during fore-aft and vertical and for the trunk translation in lateral perturbations (\Cref{Table:StatsP}). Translational peak frequencies were also significantly affected by vision (F(1) = 8.958, p = 0.010). 
Interactions between direction, segment, and average gain were significant (F(4) = 4.268, p = 0.020). Post hoc tests show that only the head peak frequencies during fore-aft perturbations are significantly affected by vision (\Cref{Table:StatsP}). Without vision, the head peak frequencies were higher (1.4 Hz on average) than with vision during fore-aft perturbations. 
Vision shows no significant effect on rotational peak gains and frequencies. The average gain between 1 and 2 Hz for the rotational response (\Cref{Table:TRdataAvg}) is significantly affected by vision (F(1) = 26.584, p $<$ 0.001). Interactions between direction, segment, and average gain are significant (F(4) = 8.509, p $<$ 0.001). Post hoc tests show that the head 1-2 Hz rotational gain of the pitch response to fore-aft and roll response to lateral are significantly affected by vision (\Cref{Table:StatsAvg}). Without vision, the 1-2 Hz head pitch gain with fore-aft perturbations increases around 18\%, while head roll increases around 25\% and trunk roll with 7\% with lateral perturbations (\Cref{Table:TRdataAvg}).    

\subsection{Head down}
Figure \ref{Figure:extra} shows body segment responses in the head-down posture. 
Peak gains are not significantly different between conditions looking forward and looking down (F(1)=4.304, p = 0.058). However the interaction with segment and direction is significant (F(4)= 6.554, p $<$ 0.001). Post hoc tests show that this is significant for the head with fore-aft perturbations and for the trunk with lateral perturbations (\Cref{Table:StatsP}). Related frequencies are significantly affected by head orientation (F(1) = 4.710, p = 0.049) with no significant interaction with segment and direction (F(4)= 2.929, p = 0.054). Figure \ref{Figure:extra} shows that particularly during the fore-aft perturbations, the peak gain was higher and shifted up around 0.5 Hz in the head-down condition compared to the reference posture (i.e., preferred posture with middle support height and looking straight forward). Peak main rotational gains are significantly affected by head orientation (F(1) = 5.112, p = 0.040). Post hoc tests show that the difference is only significant in the head main rotational response (pitch) to vertical perturbations where the peak pitch gain in the head-down condition is 65\% percent more than 'looking forward' (\Cref{Table:TRdataP}).  No significant effects on main rotational peak frequencies were found. 

\subsection{Motion amplitude}
Peak gains are higher in the condition where the amplitude of the applied vibration was scaled to 0.25 of the default signal (\Cref{Figure:extra}, \Cref{Table:TRdataP}). A repeated-measures ANOVA shows that this difference is significant (F(1) = 26.548, p $<$ 0.001) as well as the interaction with segment and direction (F(4)=9.363, p $<$ 0.001). Post hoc tests show that the difference is significant in all cases except for the pelvis during vertical and the head during lateral perturbations (\Cref{Table:StatsP}). Peak frequencies are also significantly different between low amplitude and default excitation (F(1) = 66.433, p \textless 0.001) and with significant interaction with direction and segment (F(4)=8.785, p $<$ 0.001). Low amplitude excitation signals lead to slightly higher peak gain frequencies (\Cref{Figure:extra}, \Cref{Table:TRdataP}). Peak main rotational gains are also significantly affected by motion amplitude (F(1) = 9.780, p = 0.008). Post hoc tests indicate that peak gains differ significant only for the head with fore-aft perturbations and for the trunk with lateral perturbations (\Cref{Table:StatsP}). The averaged 1-2 Hz gain is also affected by motion amplitude (\Cref{Table:TRdataAvg,Table:StatsAvg}). Similar to translational peak gains, rotational gains were higher in low amplitude conditions (\Cref{Table:TRdataP}). Rotational peak gain frequencies were also significantly affected by motion amplitude (F(1) = 6.927, p = 0.021), but with no interaction with segment and direction (F(4) = 0.420, p = 0.792). Low amplitude excitation signals led to slightly higher rotational peak gain frequencies (\Cref{Table:TRdataP}).  

\subsection{Seat pressure}
The seat Centre of Pressure (CoP) forward displacement was on average below 0.2 mm rms in all conditions and lateral displacement was also below 0.2 mm with fore-aft and vertical perturbations (Figure B.1). The lateral CoP displacement with lateral perturbations was significantly higher at 1.6 mm. Frequency domain analysis indicated a moderate coherence between lateral CoP displacement and the applied platform motion (Figure B.2). For the vertical excitation, the apparent mass was calculated by computing the transfer functions of the total seat contact force (summation of pressure signals of individual sensors) relative to the vertical acceleration of the motion platform resulting in an unrealistically low apparent mass of 15 kg with varying coherence (Figure B.3). Hence the dynamic pressure response seems underestimated calling for dynamic calibration \cite{Liu2018274} and verification measuring seat forces.

\section{Discussion} 

To achieve our scientific objectives we developed an experimental methodology to evaluate 3D vibration transmission from compliant seats to the human body. We designed wide-band motion stimuli and applied these in fore-aft, lateral and vertical direction and evaluated the translational and rotational body response in pelvis, trunk and head. Coherent kinematic results were obtained using body inertial measurements with a platform motion amplitude of only 0.3 m/s\textsuperscript{2} rms. This allowed wide-band motion (0.1-12 Hz) on a 0.7 m stroke motion platform. An exposure of 60 seconds per motion condition as in \cite{mansfield2006effect} was found sufficient to obtain coherent and consistent results from 0.34 Hz. Results below this frequency, as well as the significant effects of amplitude, will be evaluated in the time domain using 3D nonlinear models of human seat interaction \citep{MirakhorloICC2021}. Such biomechanical models can also address cross-axis nonlinearity as demonstrated by \cite{Zheng201958}. 

Frequency domain results in terms of seat to head transmissibility are comparable to previous studies. In vertical loading, transfer function gains in translation are close to one at frequencies below 4 Hz and peak gains are in the range of previously reported human body resonance frequencies (4-6 Hz) \citep{nawayseh2003non,rakheja2010biodynamics} (\Cref{Figure:Z}, \Cref{Figure:PTH}). We used analyses of variance to assess the significance of effects of posture and seat back height. Translational responses show significant effects in particular on resonances in terms of gain peak amplitude and frequency. Rotational responses show significant effects in particular at low frequencies. These effects are partially consistent for fore-aft, lateral and vertical perturbations, highlighting the added value of combined testing and statistical analysis for 3 seat motion directions. Future research shall explore contributions of translational and rotational motion to comfort perception in particular for the head. To achieve this we are currently integrating models of sensory integration \citep{Oman1982} and postural stabilization \citep{HappeeNeck2017} in full body biomechanical models \citep{MirakhorloICC2021}.


We found substantial effects of posture and seat back height on postural stabilization reflected in altered peak gains and associated frequencies in all seat motion directions. Rotational gain responses to fore-aft (pitch) and lateral (roll) were significantly affected by posture and seat back height at low frequencies (1-2 Hz). Perceived discomfort was substantially affected by posture and seat back height with the strongest discomfort being observed with a low back support with slouched posture. 
\par The low back support led to substantially lower peak gains than the middle and high support during fore-aft and vertical perturbations (\Cref{Table:TRdataP}) in particular for head rotation (\Cref{Figure:PTH}). Low frequency (1-2 Hz) gains were significantly lower with low back support during lateral perturbations. We attribute these findings to the constraining effect of the back support on trunk motion. As outlined in the introduction we expected larger head motions with more support. This expectation was based on tests with rigid high back support \citep{forbes2013dependency} and without back support \citep{van2016trunk} and is now confirmed with compliant back support. Presumably, the additional motion freedom of the thorax and lumbar spine with low support allows for more effective head-in-space stabilization. However, the low support is also rated as least comfortable, in particular with the slouched posture, as discussed further below. Hence, the search for more comfortable car seats could explore seat backs that support against gravity and vehicle motion induced loading but which do not so much constrain upper back motion. In line with our findings, it was shown that an arm support can constrain trunk motion but leads to higher head translational motions in response to multiple axes perturbations \citep{rahmatalla2010quasi}. The vertical STHT has been studied comparing conditions with and without back support  \citep{toward2011transmission} reporting no effect on peak gains in line with our findings (\Cref{Table:StatsP}). 

Participants rated the slouched sitting posture more discomforting than the preferred and erect postures. Potential discomforting stimuli include 1) back support and seat pressure, 2) body posture (e.g., high stress in joint structures due to uncomfortable joint orientations approaching the range of motion, and 3) body motion. Regarding point 1, the experimental seat was presumably overall less comfortable than commercial car seats due to higher peak pressures associated with the limited back support surface. This effect may have been most pronounced in the slouched postures and/or with low back support. Regarding point 2, the slouched posture itself may also be perceived as less comfortable, as reported in studies on office chairs \citep{Vergara2000} and train seats \citep{Groenesteijn2014}. Regarding point 3, it has been shown that discomfort can be predicted by the acceleration profiles of the seat, back, and feet \citep{Basri2013}. In our study slouched leads to higher translational peak gains than preferred and erect postures. Interestingly, the perceived trunk ratings followed the same pattern as the overall discomfort ratings across conditions (Figure  \ref{perc_support}). This indicates that the trunk support and the resulting trunk motion are partly responsible for the experienced discomfort. This highlights the need to assess trunk related comfort metrics and not only use head motion, as head motion was actually reduced in the least comfortable condition with low  backrest and slouched posture.


\par A reduced perturbation signal magnitude resulted in increase main response peak gains in all perturbation directions for head and trunk, accompanied with higher peak frequencies (\Cref{Table:TRdataP}). Previous studies found  similar non-linear effects of seat vibration magnitude \citep{bhiwapurkar2016effects,bhiwapurkar2019effects,nawayseh2003non}. These findings were explained by non-linear muscle yielding in response to increasing motion magnitudes \citep{nawayseh2003non,matsumoto2002effect}. In the arm a similar yielding was found in relax tasks while position tasks elicited stiffening with higher perturbation amplitude associated with increased muscle activity \citep{happee2015nonlinear}.

The eyes closed (EC) condition was tested to support modelling of vestibular and visual contributions to postural stabilization. Exposure to vibration without vision increased head peak gains with no clear effect on the pelvis and trunk (\Cref{Table:TRdataP,Table:StatsP}). The effect of vision is particularly evident for rotational gains at low frequencies (1-2 Hz) during fore-aft and lateral perturbations, which can be explained by a more dominant effect of visual feedback on postural stabilization at lower frequencies \citep{van2016trunk,forbes2013dependency}. 

In this study, we also asked the participants to sit in a head down posture that mimics working on a handheld tablet or smartphone. As automated driving provides the opportunity to perform non-driving tasks instead of paying attention to the traffic or the road, this sitting posture might be very common in the future. The flexed orientation of the head changed the dynamics profoundly as the head started resonating more when vibrated in the fore-aft (i.e., linear acceleration, Fig. \ref{Figure:extra}a) and vertical direction (i.e., pitch angular acceleration, Fig. \ref{Figure:extra}b). Higher averaged low frequency (1-2 Hz) gains were also found for the head down posture (\Cref{Table:TRdataAvg}). This result concurs with fore-aft vibration experiments where head down postures elicited increased head motion and discomfort \citep{rahmatalla2011effective}. Looking forward at an auxiliary display, rather than looking down was also shown to reduce car sickness while driving a slalom, where beneficial effects were associated with peripheral outside vision \citep{kuiper2019moving}.   

\par In this study we present human body responses interacting with a compliant seat. It shall be noted that our results will be affected by the actual seat compliance as well as the absence of seat back wings. Hands were placed on the lap which can dampen the higher modes of vibration \citep{matsumoto1998movement} but effects will be limited as we studied frequencies below 12 Hz. Future modelling studies will address contributions of the seat and the human body in vibration transmission.   

\section{Conclusion}
Our experimental methodology revealed significant effects of experimental conditions on body kinematics which were partially consistent across seat motion directions. Seat back support height and sitting posture affect trunk and head postural stabilization in all motion directions with a more evident effect in fore-aft and vertical. Low STHT gains for low back support confirmed our hypothesis of its advantage for head stabilization. The head-down posture caused higher head fore-aft and pitch responses. Reducing the seat motion amplitude resulted in higher peak gains and frequencies. Without vision, low frequency (1-2 Hz) head rotation increased in pitch with fore-aft perturbations and in roll with lateral perturbations. The collected human response data will support the development of human models capturing postural stabilization and predicting comfort in dynamic interaction with compliant seats.

\section{Acknowledgement} 
We acknowledge the support of Toyota Motor Corporation. The EMG analysis was performed by Anna Marbus and Marko Cvetković contributed to the statistical analysis.

 \bibliographystyle{cas-model2-names}

 \bibliography{cas-refs}

\begin{thebibliography}{52}
\expandafter\ifx\csname natexlab\endcsname\relax\def\natexlab#1{#1}\fi
\providecommand{\url}[1]{\texttt{#1}}
\providecommand{\href}[2]{#2}
\providecommand{\path}[1]{#1}
\providecommand{\DOIprefix}{doi:}
\providecommand{\ArXivprefix}{arXiv:}
\providecommand{\URLprefix}{URL: }
\providecommand{\Pubmedprefix}{pmid:}
\providecommand{\doi}[1]{\href{http://dx.doi.org/#1}{\path{#1}}}
\providecommand{\Pubmed}[1]{\href{pmid:#1}{\path{#1}}}
\providecommand{\bibinfo}[2]{#2}
\ifx\xfnm\relax \def\xfnm[#1]{\unskip,\space#1}\fi
\bibitem[{Adam et~al.(2020)Adam, {Abdul Jalil}, {Md. Rezali} and
  Ng}]{ADAM2020103014}
\bibinfo{author}{Adam, S.}, \bibinfo{author}{{Abdul Jalil}, N.},
  \bibinfo{author}{{Md. Rezali}, K.}, \bibinfo{author}{Ng, Y.},
  \bibinfo{year}{2020}.
\newblock \bibinfo{title}{The effect of posture and vibration magnitude on the
  vertical vibration transmissibility of tractor suspension system}.
\newblock \bibinfo{journal}{International Journal of Industrial Ergonomics}
  \bibinfo{volume}{80}, \bibinfo{pages}{103014}.
\newblock \URLprefix
  \url{https://www.sciencedirect.com/science/article/pii/S0169814119305396},
  \DOIprefix\doi{https://doi.org/10.1016/j.ergon.2020.103014}.
\bibitem[{Basri and Griffin(2013)}]{Basri2013}
\bibinfo{author}{Basri, B.}, \bibinfo{author}{Griffin, M.J.},
  \bibinfo{year}{2013}.
\newblock \bibinfo{title}{Predicting discomfort from whole-body vertical
  vibration when sitting with an inclined backrest.}
\newblock \bibinfo{journal}{Appl Ergon} \bibinfo{volume}{44},
  \bibinfo{pages}{423--434}.
\newblock \DOIprefix\doi{10.1016/j.apergo.2012.10.006}.
\bibitem[{Basri and Griffin(2014)}]{basri2014application}
\bibinfo{author}{Basri, B.}, \bibinfo{author}{Griffin, M.J.},
  \bibinfo{year}{2014}.
\newblock \bibinfo{title}{The application of seat values for predicting how
  compliant seats with backrests influence vibration discomfort}.
\newblock \bibinfo{journal}{Applied ergonomics} \bibinfo{volume}{45},
  \bibinfo{pages}{1461--1474}.
\bibitem[{Bhiwapurkar et~al.(2019)Bhiwapurkar, Saran and
  Harsha}]{bhiwapurkar2019effects}
\bibinfo{author}{Bhiwapurkar, M.}, \bibinfo{author}{Saran, V.},
  \bibinfo{author}{Harsha, S.}, \bibinfo{year}{2019}.
\newblock \bibinfo{title}{Effects of posture and vibration magnitude on seat to
  head transmissibility during exposure to fore-and-aft vibration}.
\newblock \bibinfo{journal}{Journal of Low Frequency Noise, Vibration and
  Active Control} \bibinfo{volume}{38}, \bibinfo{pages}{826--838}.
\bibitem[{Bhiwapurkar et~al.(2016)Bhiwapurkar, Saran and
  Harsha}]{bhiwapurkar2016effects}
\bibinfo{author}{Bhiwapurkar, M.K.}, \bibinfo{author}{Saran, V.},
  \bibinfo{author}{Harsha, S.}, \bibinfo{year}{2016}.
\newblock \bibinfo{title}{Effects of vibration magnitude and posture on
  seat-to-head-transmissibility responses of seated occupants exposed to
  lateral vibration}.
\newblock \bibinfo{journal}{International Journal of Vehicle Noise and
  Vibration} \bibinfo{volume}{12}, \bibinfo{pages}{42--59}.
\bibitem[{Boileau and Rakheja(1998)}]{boileau1998whole}
\bibinfo{author}{Boileau, P.{\'E}.}, \bibinfo{author}{Rakheja, S.},
  \bibinfo{year}{1998}.
\newblock \bibinfo{title}{Whole-body vertical biodynamic response
  characteristics of the seated vehicle driver: measurement and model
  development}.
\newblock \bibinfo{journal}{International Journal of Industrial Ergonomics}
  \bibinfo{volume}{22}, \bibinfo{pages}{449--472}.
\bibitem[{Corbridge and Griffin(1986)}]{corbridge1986vibration}
\bibinfo{author}{Corbridge, C.}, \bibinfo{author}{Griffin, M.},
  \bibinfo{year}{1986}.
\newblock \bibinfo{title}{Vibration and comfort: vertical and lateral motion in
  the range 0{\textperiodcentered} 5 to 5{\textperiodcentered} 0 hz}.
\newblock \bibinfo{journal}{Ergonomics} \bibinfo{volume}{29},
  \bibinfo{pages}{249--272}.
\bibitem[{Dialynas et~al.(2019)Dialynas, Haan, Schouten, Happee and
  Schwab}]{dialynas2019}
\bibinfo{author}{Dialynas, G.}, \bibinfo{author}{Haan, J.},
  \bibinfo{author}{Schouten, A.}, \bibinfo{author}{Happee, R.},
  \bibinfo{author}{Schwab, A.}, \bibinfo{year}{2019}.
\newblock \bibinfo{title}{The dynamic response of the bicycle rider’s body to
  vertical, fore-and-aft, and lateral perturbations}.
\newblock \bibinfo{journal}{Proceedings of the Institution of Mechanical
  Engineers, Part D: Journal of Automobile Engineering} \bibinfo{volume}{234},
  \bibinfo{pages}{095440701989128}.
\newblock \DOIprefix\doi{10.1177/0954407019891289}.
\bibitem[{Dong et~al.(2019)Dong, He, Du, Cao and Huang}]{dong2019effect}
\bibinfo{author}{Dong, R.c.}, \bibinfo{author}{He, L.}, \bibinfo{author}{Du,
  W.}, \bibinfo{author}{Cao, Z.k.}, \bibinfo{author}{Huang, Z.l.},
  \bibinfo{year}{2019}.
\newblock \bibinfo{title}{Effect of sitting posture and seat on biodynamic
  responses of internal human body simulated by finite element modeling of
  body-seat system}.
\newblock \bibinfo{journal}{Journal of Sound and Vibration}
  \bibinfo{volume}{438}, \bibinfo{pages}{543--554}.
\bibitem[{van Drunen et~al.(2016)van Drunen, van~der Helm, van Die{\"e}n and
  Happee}]{van2016trunk}
\bibinfo{author}{van Drunen, P.}, \bibinfo{author}{van~der Helm, F.C.},
  \bibinfo{author}{van Die{\"e}n, J.H.}, \bibinfo{author}{Happee, R.},
  \bibinfo{year}{2016}.
\newblock \bibinfo{title}{Trunk stabilization during sagittal pelvic tilt: from
  trunk-on-pelvis to trunk-in-space due to vestibular and visual feedback}.
\newblock \bibinfo{journal}{Journal of neurophysiology} \bibinfo{volume}{115},
  \bibinfo{pages}{1381--1388}.
\bibitem[{Fairley and Griffin(1989)}]{fairley1989apparent}
\bibinfo{author}{Fairley, T.E.}, \bibinfo{author}{Griffin, M.J.},
  \bibinfo{year}{1989}.
\newblock \bibinfo{title}{The apparent mass of the seated human body: vertical
  vibration}.
\newblock \bibinfo{journal}{Journal of Biomechanics} \bibinfo{volume}{22},
  \bibinfo{pages}{81--94}.
\bibitem[{Forbes et~al.(2013)Forbes, de~Bruijn, Schouten, van~der Helm and
  Happee}]{forbes2013dependency}
\bibinfo{author}{Forbes, P.A.}, \bibinfo{author}{de~Bruijn, E.},
  \bibinfo{author}{Schouten, A.C.}, \bibinfo{author}{van~der Helm, F.C.},
  \bibinfo{author}{Happee, R.}, \bibinfo{year}{2013}.
\newblock \bibinfo{title}{Dependency of human neck reflex responses on the
  bandwidth of pseudorandom anterior-posterior torso perturbations}.
\newblock \bibinfo{journal}{Experimental brain research} \bibinfo{volume}{226},
  \bibinfo{pages}{1--14}.
\bibitem[{Frey et~al.(2020)Frey, Greene and {De Carvalho}}]{frey2020}
\bibinfo{author}{Frey, M.}, \bibinfo{author}{Greene, R.}, \bibinfo{author}{{De
  Carvalho}, D.}, \bibinfo{year}{2020}.
\newblock \bibinfo{title}{What is the best way to collect maximum forward
  lumbar spine flexion values for normalizing posture to range of motion?}
\newblock \bibinfo{journal}{Journal of Biomechanics} \bibinfo{volume}{103},
  \bibinfo{pages}{109706}.
\newblock \URLprefix
  \url{http://www.sciencedirect.com/science/article/pii/S0021929020301226},
  \DOIprefix\doi{https://doi.org/10.1016/j.jbiomech.2020.109706}.
\bibitem[{Groenesteijn et~al.(2014)Groenesteijn, van Mastrigt, Gallais, Blok,
  Kuijt-Evers and Vink}]{Groenesteijn2014}
\bibinfo{author}{Groenesteijn, L.}, \bibinfo{author}{van Mastrigt, S.H.},
  \bibinfo{author}{Gallais, C.}, \bibinfo{author}{Blok, M.},
  \bibinfo{author}{Kuijt-Evers, L.}, \bibinfo{author}{Vink, P.},
  \bibinfo{year}{2014}.
\newblock \bibinfo{title}{Activities, postures and comfort perception of train
  passengers as input for train seat design}.
\newblock \bibinfo{journal}{Ergonomics} \bibinfo{volume}{57},
  \bibinfo{pages}{1154--1165}.
\newblock \URLprefix \url{https://doi.org/10.1080/00140139.2014.914577},
  \DOIprefix\doi{10.1080/00140139.2014.914577},
  \href{http://arxiv.org/abs/https://doi.org/10.1080/00140139.2014.914577}{\tt
  arXiv:https://doi.org/10.1080/00140139.2014.914577}. \bibinfo{note}{pMID:
  24831434}.
\bibitem[{Happee et~al.(2017)Happee, de~Bruijn, Forbes and van~der
  Helm}]{HappeeNeck2017}
\bibinfo{author}{Happee, R.}, \bibinfo{author}{de~Bruijn, E.},
  \bibinfo{author}{Forbes, P.}, \bibinfo{author}{van~der Helm, F.},
  \bibinfo{year}{2017}.
\newblock \bibinfo{title}{Dynamic head-neck stabilization and modulation with
  perturbation bandwidth investigated using a multisegment neuromuscular
  model}.
\newblock \bibinfo{journal}{J Biomechanics} .
\bibitem[{Happee et~al.(2015)Happee, de~Vlugt and van
  Vliet}]{happee2015nonlinear}
\bibinfo{author}{Happee, R.}, \bibinfo{author}{de~Vlugt, E.},
  \bibinfo{author}{van Vliet, B.}, \bibinfo{year}{2015}.
\newblock \bibinfo{title}{Nonlinear 2d arm dynamics in response to continuous
  and pulse‑shaped force perturbations}.
\newblock \bibinfo{journal}{Experimental brain research} \bibinfo{volume}{233},
  \bibinfo{pages}{39--52}.
\bibitem[{Jalil and Griffin(2007a)}]{jalil2007fore}
\bibinfo{author}{Jalil, N.A.A.}, \bibinfo{author}{Griffin, M.J.},
  \bibinfo{year}{2007}a.
\newblock \bibinfo{title}{Fore-and-aft transmissibility of backrests: Effect of
  backrest inclination, seat-pan inclination, and measurement location}.
\newblock \bibinfo{journal}{Journal of sound and vibration}
  \bibinfo{volume}{299}, \bibinfo{pages}{99--108}.
\bibitem[{Jalil and Griffin(2007b)}]{Jalil2007}
\bibinfo{author}{Jalil, N.A.A.}, \bibinfo{author}{Griffin, M.J.},
  \bibinfo{year}{2007}b.
\newblock \bibinfo{title}{Fore-and-aft transmissibility of backrests: Variation
  with height above the seat surface and non-linearity}.
\newblock \bibinfo{journal}{Journal of Sound and Vibration}
  \bibinfo{volume}{299}, \bibinfo{pages}{109 -- 122}.
\newblock \URLprefix
  \url{http://www.sciencedirect.com/science/article/pii/S0022460X06005657},
  \DOIprefix\doi{https://doi.org/10.1016/j.jsv.2006.06.057}.
\bibitem[{Khusro et~al.(2020)Khusro, Zheng, Grottoli and
  Shyrokau}]{khusro2020mpc}
\bibinfo{author}{Khusro, Y.R.}, \bibinfo{author}{Zheng, Y.},
  \bibinfo{author}{Grottoli, M.}, \bibinfo{author}{Shyrokau, B.},
  \bibinfo{year}{2020}.
\newblock \bibinfo{title}{Mpc-based motion-cueing algorithm for a 6-dof driving
  simulator with actuator constraints}.
\newblock \bibinfo{journal}{Vehicles} \bibinfo{volume}{2},
  \bibinfo{pages}{625--647}.
\bibitem[{Kuiper et~al.(2019)Kuiper, Bos, Diels and
  Cammaerts}]{kuiper2019moving}
\bibinfo{author}{Kuiper, O.X.}, \bibinfo{author}{Bos, J.E.},
  \bibinfo{author}{Diels, C.}, \bibinfo{author}{Cammaerts, K.},
  \bibinfo{year}{2019}.
\newblock \bibinfo{title}{Moving base driving simulators’ potential for
  carsickness research}.
\newblock \bibinfo{journal}{Applied ergonomics} \bibinfo{volume}{81},
  \bibinfo{pages}{102889}.
\bibitem[{Kyriakidis et~al.(2015)Kyriakidis, Happee and
  de~Winter}]{kyriakidis2015public}
\bibinfo{author}{Kyriakidis, M.}, \bibinfo{author}{Happee, R.},
  \bibinfo{author}{de~Winter, J.C.}, \bibinfo{year}{2015}.
\newblock \bibinfo{title}{Public opinion on automated driving: Results of an
  international questionnaire among 5000 respondents}.
\newblock \bibinfo{journal}{Transportation research part F: traffic psychology
  and behaviour} \bibinfo{volume}{32}, \bibinfo{pages}{127--140}.
\bibitem[{Li and Huang(2020)}]{Li2020}
\bibinfo{author}{Li, J.}, \bibinfo{author}{Huang, Y.}, \bibinfo{year}{2020}.
\newblock \bibinfo{title}{The effects of the duration on the subjective
  discomfort of a rigid seat and a cushioned automobile seat}.
\newblock \bibinfo{journal}{International Journal of Industrial Ergonomics}
  \bibinfo{volume}{79}.
\newblock \URLprefix
  \url{https://www.scopus.com/inward/record.uri?eid=2-s2.0-85089475250&doi=10.1016%2fj.ergon.2020.103007&partnerID=40&md5=6675f4a30f963cd6218d3e0657d521d1},
  \DOIprefix\doi{10.1016/j.ergon.2020.103007}.
\bibitem[{Liu and Griffin(2018)}]{Liu2018274}
\bibinfo{author}{Liu, C.}, \bibinfo{author}{Griffin, M.}, \bibinfo{year}{2018}.
\newblock \bibinfo{title}{Measuring vibration-induced variations in pressures
  between the human body and a seat}.
\newblock \bibinfo{journal}{International Journal of Industrial Ergonomics}
  \bibinfo{volume}{67}, \bibinfo{pages}{274--282}.
\newblock \URLprefix
  \url{https://www.scopus.com/inward/record.uri?eid=2-s2.0-85049024351&doi=10.1016%2fj.ergon.2018.05.006&partnerID=40&md5=afdaeae660f2896df68c791539305245},
  \DOIprefix\doi{10.1016/j.ergon.2018.05.006}.
\bibitem[{Mandapuram et~al.(2012)Mandapuram, Rakheja, Boileau and
  Maeda}]{mandapuram2012apparent}
\bibinfo{author}{Mandapuram, S.}, \bibinfo{author}{Rakheja, S.},
  \bibinfo{author}{Boileau, P.{\'E}.}, \bibinfo{author}{Maeda, S.},
  \bibinfo{year}{2012}.
\newblock \bibinfo{title}{Apparent mass and head vibration transmission
  responses of seated body to three translational axis vibration}.
\newblock \bibinfo{journal}{International Journal of Industrial Ergonomics}
  \bibinfo{volume}{42}, \bibinfo{pages}{268--277}.
\bibitem[{Mansfield et~al.(2006)Mansfield, Holmlund, Lundstr{\"o}m, Lenzuni and
  Nataletti}]{mansfield2006effect}
\bibinfo{author}{Mansfield, N.}, \bibinfo{author}{Holmlund, P.},
  \bibinfo{author}{Lundstr{\"o}m, R.}, \bibinfo{author}{Lenzuni, P.},
  \bibinfo{author}{Nataletti, P.}, \bibinfo{year}{2006}.
\newblock \bibinfo{title}{Effect of vibration magnitude, vibration spectrum and
  muscle tension on apparent mass and cross axis transfer functions during
  whole-body vibration exposure}.
\newblock \bibinfo{journal}{Journal of biomechanics} \bibinfo{volume}{39},
  \bibinfo{pages}{3062--3070}.
\bibitem[{Matsumoto and Griffin(1998)}]{matsumoto1998movement}
\bibinfo{author}{Matsumoto, Y.}, \bibinfo{author}{Griffin, M.},
  \bibinfo{year}{1998}.
\newblock \bibinfo{title}{Movement of the upper-body of seated subjects exposed
  to vertical whole-body vibration at the principal resonance frequency}.
\newblock \bibinfo{journal}{Journal of Sound and Vibration}
  \bibinfo{volume}{215}, \bibinfo{pages}{743--762}.
\bibitem[{Matsumoto and Griffin(2002)}]{matsumoto2002effect}
\bibinfo{author}{Matsumoto, Y.}, \bibinfo{author}{Griffin, M.},
  \bibinfo{year}{2002}.
\newblock \bibinfo{title}{Effect of muscle tension on non-linearities in the
  apparent masses of seated subjects exposed to vertical whole-body vibration}.
\newblock \bibinfo{journal}{Journal of Sound and Vibration}
  \bibinfo{volume}{253}, \bibinfo{pages}{77--92}.
\bibitem[{Mirakhorlo et~al.(2021)Mirakhorlo, Kluft, Irmak, Shyrokau and
  Happee}]{MirakhorloICC2021}
\bibinfo{author}{Mirakhorlo, M.}, \bibinfo{author}{Kluft, N.},
  \bibinfo{author}{Irmak, T.}, \bibinfo{author}{Shyrokau, B.},
  \bibinfo{author}{Happee, R.}, \bibinfo{year}{2021}.
\newblock \bibinfo{title}{Simulating 3d human postural stabilization in
  vibration and dynamic driving}.
\newblock \bibinfo{journal}{International Comfort Conference} .
\bibitem[{Nawayseh(2015)}]{Nawayseh201582}
\bibinfo{author}{Nawayseh, N.}, \bibinfo{year}{2015}.
\newblock \bibinfo{title}{Effect of the seating condition on the transmission
  of vibration through the seat pan and backrest}.
\newblock \bibinfo{journal}{International Journal of Industrial Ergonomics}
  \bibinfo{volume}{45}, \bibinfo{pages}{82--90}.
\newblock \URLprefix
  \url{https://www.scopus.com/inward/record.uri?eid=2-s2.0-84921307979&doi=10.1016%2fj.ergon.2014.12.005&partnerID=40&md5=4bbe6d82af1441d4cc8b0c05aa4f1512},
  \DOIprefix\doi{10.1016/j.ergon.2014.12.005}.
\bibitem[{Nawayseh et~al.(2020)Nawayseh, Alchakouch and
  Hamdan}]{nawayseh2020tri}
\bibinfo{author}{Nawayseh, N.}, \bibinfo{author}{Alchakouch, A.},
  \bibinfo{author}{Hamdan, S.}, \bibinfo{year}{2020}.
\newblock \bibinfo{title}{Tri-axial transmissibility to the head and spine of
  seated human subjects exposed to fore-and-aft whole-body vibration}.
\newblock \bibinfo{journal}{Journal of Biomechanics} \bibinfo{volume}{109},
  \bibinfo{pages}{109927}.
\bibitem[{Nawayseh and Griffin(2003)}]{nawayseh2003non}
\bibinfo{author}{Nawayseh, N.}, \bibinfo{author}{Griffin, M.},
  \bibinfo{year}{2003}.
\newblock \bibinfo{title}{Non-linear dual-axis biodynamic response to vertical
  whole-body vibration}.
\newblock \bibinfo{journal}{Journal of Sound and Vibration}
  \bibinfo{volume}{268}, \bibinfo{pages}{503--523}.
\bibitem[{Nawayseh and Griffin(2005)}]{nawayseh2005non}
\bibinfo{author}{Nawayseh, N.}, \bibinfo{author}{Griffin, M.},
  \bibinfo{year}{2005}.
\newblock \bibinfo{title}{Non-linear dual-axis biodynamic response to
  fore-and-aft whole-body vibration}.
\newblock \bibinfo{journal}{Journal of Sound and Vibration}
  \bibinfo{volume}{282}, \bibinfo{pages}{831--862}.
\bibitem[{Oman(1982)}]{Oman1982}
\bibinfo{author}{Oman, C.M.}, \bibinfo{year}{1982}.
\newblock \bibinfo{title}{{A heuristic mathematical model for the dynamics of
  sensory conflict and motion sickness.}}
\newblock \bibinfo{journal}{Acta Oto-Laryngologica} \bibinfo{volume}{94},
  \bibinfo{pages}{4--44}.
\newblock \URLprefix \url{https:/doi.org/10.3109/00016488209108197}.
\bibitem[{Paddan and Griffin(1998)}]{paddan1998review}
\bibinfo{author}{Paddan, G.}, \bibinfo{author}{Griffin, M.},
  \bibinfo{year}{1998}.
\newblock \bibinfo{title}{A review of the transmission of translational seat
  vibration to the head}.
\newblock \bibinfo{journal}{Journal of Sound and Vibration}
  \bibinfo{volume}{215}, \bibinfo{pages}{863--882}.
\bibitem[{Papaioannou et~al.(2021)Papaioannou, Jerrelind, Drugge and
  Shyrokau}]{papaioannou2021assessment}
\bibinfo{author}{Papaioannou, G.}, \bibinfo{author}{Jerrelind, J.},
  \bibinfo{author}{Drugge, L.}, \bibinfo{author}{Shyrokau, B.},
  \bibinfo{year}{2021}.
\newblock \bibinfo{title}{Assessment of optimal passive suspensions regarding
  motion sickness mitigation in different road profiles and sitting
  conditions}.
\newblock \bibinfo{journal}{IEEE International Intelligent Transportation
  Systems Conference} .
\bibitem[{Rahmatalla and DeShaw(2011)}]{rahmatalla2011effective}
\bibinfo{author}{Rahmatalla, S.}, \bibinfo{author}{DeShaw, J.},
  \bibinfo{year}{2011}.
\newblock \bibinfo{title}{Effective seat-to-head transmissibility in whole-body
  vibration: Effects of posture and arm position}.
\newblock \bibinfo{journal}{Journal of Sound and Vibration}
  \bibinfo{volume}{330}, \bibinfo{pages}{6277--6286}.
\bibitem[{Rahmatalla et~al.(2010)Rahmatalla, Smith, Meusch, Xia, Marler and
  Contratto}]{rahmatalla2010quasi}
\bibinfo{author}{Rahmatalla, S.}, \bibinfo{author}{Smith, R.},
  \bibinfo{author}{Meusch, J.}, \bibinfo{author}{Xia, T.},
  \bibinfo{author}{Marler, T.}, \bibinfo{author}{Contratto, M.},
  \bibinfo{year}{2010}.
\newblock \bibinfo{title}{A quasi-static discomfort measure in whole-body
  vibration}.
\newblock \bibinfo{journal}{Industrial health} \bibinfo{volume}{48},
  \bibinfo{pages}{645--653}.
\bibitem[{Rakheja et~al.(2010)Rakheja, Dong, Patra, Boileau, Marcotte and
  Warren}]{rakheja2010biodynamics}
\bibinfo{author}{Rakheja, S.}, \bibinfo{author}{Dong, R.},
  \bibinfo{author}{Patra, S.}, \bibinfo{author}{Boileau, P.{\'E}.},
  \bibinfo{author}{Marcotte, P.}, \bibinfo{author}{Warren, C.},
  \bibinfo{year}{2010}.
\newblock \bibinfo{title}{Biodynamics of the human body under whole-body
  vibration: Synthesis of the reported data}.
\newblock \bibinfo{journal}{International Journal of Industrial Ergonomics}
  \bibinfo{volume}{40}, \bibinfo{pages}{710--732}.
\bibitem[{Reason and Brand(1975)}]{Reason1975}
\bibinfo{author}{Reason, J.T.}, \bibinfo{author}{Brand, J.J.},
  \bibinfo{year}{1975}.
\newblock \bibinfo{title}{Motion sickness.}
\newblock \bibinfo{publisher}{Academic Press}, \bibinfo{address}{Oxford,
  England}.
\bibitem[{Schepers et~al.(2018)Schepers, Giuberti and Bellusci}]{white2018}
\bibinfo{author}{Schepers, M.}, \bibinfo{author}{Giuberti, M.},
  \bibinfo{author}{Bellusci, G.}, \bibinfo{year}{2018}.
\newblock \bibinfo{title}{Xsens mvn: Consistent tracking of human motion using
  inertial sensing} \DOIprefix\doi{10.13140/RG.2.2.22099.07205}.
\bibitem[{Song et~al.(2017)Song, Ahn, Jeong and Yoo}]{song2017subjective}
\bibinfo{author}{Song, J.}, \bibinfo{author}{Ahn, S.}, \bibinfo{author}{Jeong,
  W.}, \bibinfo{author}{Yoo, W.}, \bibinfo{year}{2017}.
\newblock \bibinfo{title}{Subjective absolute discomfort threshold due to idle
  vibration in passenger vehicles according to sitting posture}.
\newblock \bibinfo{journal}{International Journal of Automotive Technology}
  \bibinfo{volume}{18}, \bibinfo{pages}{293--300}.
\bibitem[{Tiemessen et~al.(2007)Tiemessen, Hulshof and
  Frings-Dresen}]{tiemessen2007overview}
\bibinfo{author}{Tiemessen, I.J.}, \bibinfo{author}{Hulshof, C.T.},
  \bibinfo{author}{Frings-Dresen, M.H.}, \bibinfo{year}{2007}.
\newblock \bibinfo{title}{An overview of strategies to reduce whole-body
  vibration exposure on drivers: A systematic review}.
\newblock \bibinfo{journal}{International Journal of Industrial Ergonomics}
  \bibinfo{volume}{37}, \bibinfo{pages}{245--256}.
\bibitem[{Toward and Griffin(2009)}]{toward2009apparent}
\bibinfo{author}{Toward, M.G.}, \bibinfo{author}{Griffin, M.J.},
  \bibinfo{year}{2009}.
\newblock \bibinfo{title}{Apparent mass of the human body in the vertical
  direction: Effect of seat backrest}.
\newblock \bibinfo{journal}{Journal of Sound and Vibration}
  \bibinfo{volume}{327}, \bibinfo{pages}{657--669}.
\bibitem[{Toward and Griffin(2011)}]{toward2011transmission}
\bibinfo{author}{Toward, M.G.}, \bibinfo{author}{Griffin, M.J.},
  \bibinfo{year}{2011}.
\newblock \bibinfo{title}{The transmission of vertical vibration through seats:
  Influence of the characteristics of the human body}.
\newblock \bibinfo{journal}{Journal of Sound and Vibration}
  \bibinfo{volume}{330}, \bibinfo{pages}{6526--6543}.
\bibitem[{van Veen et~al.(2015)van Veen, Orlinskiy, Franz and
  Vink}]{vanVeen2015}
\bibinfo{author}{van Veen, S.}, \bibinfo{author}{Orlinskiy, V.},
  \bibinfo{author}{Franz, M.}, \bibinfo{author}{Vink, P.},
  \bibinfo{year}{2015}.
\newblock \bibinfo{title}{Investigating car passenger well-being related to a
  seat imposing continuous posture variation}.
\newblock \bibinfo{journal}{Journal of Ergonomics 5 (3) 2015} .
\bibitem[{Vergara and Page(2000)}]{Vergara2000}
\bibinfo{author}{Vergara, M.}, \bibinfo{author}{Page, {\'A}.},
  \bibinfo{year}{2000}.
\newblock \bibinfo{title}{System to measure the use of the backrest in
  sitting-posture office tasks}.
\newblock \bibinfo{journal}{Applied Ergonomics} \bibinfo{volume}{31},
  \bibinfo{pages}{247--254}.
\newblock \URLprefix
  \url{https://www.sciencedirect.com/science/article/pii/S0003687099000563},
  \DOIprefix\doi{https://doi.org/10.1016/S0003-6870(99)00056-3}.
\bibitem[{Wu and Qiu(2020)}]{JunWuRigid}
\bibinfo{author}{Wu, Y.}, \bibinfo{author}{Qiu, Y.}, \bibinfo{year}{2020}.
\newblock \bibinfo{title}{Modelling of seated human body exposed to combined
  vertical, lateral and roll vibrations}.
\newblock \bibinfo{journal}{Journal of Sound and Vibration}
  \bibinfo{volume}{485}, \bibinfo{pages}{115509}.
\newblock \DOIprefix\doi{https://doi.org/10.1016/j.jsv.2020.115509}.
\bibitem[{Wu and Qiu(2021)}]{JunWuTrain}
\bibinfo{author}{Wu, Y.}, \bibinfo{author}{Qiu, Y.}, \bibinfo{year}{2021}.
\newblock \bibinfo{title}{Modeling and analysis of a train seat with occupant
  exposed to combined lateral, vertical and roll vibration}.
\newblock \bibinfo{journal}{Journal of Sound and Vibration}
  \bibinfo{volume}{496}, \bibinfo{pages}{115920}.
\newblock \DOIprefix\doi{https://doi.org/10.1016/j.jsv.2020.115920}.
\bibitem[{Zatsiorsky(2002)}]{Zatsiorsky2002}
\bibinfo{author}{Zatsiorsky, V.M.}, \bibinfo{year}{2002}.
\newblock \bibinfo{title}{Kinetics of human motion}.
\newblock \bibinfo{publisher}{Human Kinetics}.
\bibitem[{Zhang et~al.(2015)Zhang, Qiu and Griffin}]{zhang2015transmission}
\bibinfo{author}{Zhang, X.}, \bibinfo{author}{Qiu, Y.},
  \bibinfo{author}{Griffin, M.J.}, \bibinfo{year}{2015}.
\newblock \bibinfo{title}{Transmission of vertical vibration through a seat:
  Effect of thickness of foam cushions at the seat pan and the backrest}.
\newblock \bibinfo{journal}{International Journal of Industrial Ergonomics}
  \bibinfo{volume}{48}, \bibinfo{pages}{36--45}.
\bibitem[{Zheng et~al.(2019)Zheng, Qiu and Griffin}]{Zheng201958}
\bibinfo{author}{Zheng, G.}, \bibinfo{author}{Qiu, Y.},
  \bibinfo{author}{Griffin, M.}, \bibinfo{year}{2019}.
\newblock \bibinfo{title}{Fore-and-aft and dual-axis vibration of the seated
  human body: Nonlinearity, cross-axis coupling, and associations between
  resonances in the transmissibility and apparent mass}.
\newblock \bibinfo{journal}{International Journal of Industrial Ergonomics}
  \bibinfo{volume}{69}, \bibinfo{pages}{58--65}.
\newblock \URLprefix
  \url{https://www.scopus.com/inward/record.uri?eid=2-s2.0-85055338771&doi=10.1016%2fj.ergon.2018.08.007&partnerID=40&md5=fddfb39ecd83a3da73b85dd889b849c3},
  \DOIprefix\doi{10.1016/j.ergon.2018.08.007}.
\bibitem[{Zheng et~al.(2021)Zheng, Shyrokau and Keviczky}]{zheng2021comfort}
\bibinfo{author}{Zheng, Y.}, \bibinfo{author}{Shyrokau, B.},
  \bibinfo{author}{Keviczky, T.}, \bibinfo{year}{2021}.
\newblock \bibinfo{title}{Comfort and time efficiency: A roundabout case
  study}.
\newblock \bibinfo{journal}{IEEE International Intelligent Transportation
  Systems Conference} .

\end{thebibliography}
\newpage 
\appendix


APPENDICES TO BE PUBLISHED ON-LINE ONLY!!!
\section{Participant details, posture and Range of Motion}
\label{sec:sample:descriptives}

\subsection{Participant descriptives}
\renewcommand{\thefigure}{A.\arabic{figure}}

\setcounter{table}{0}
\renewcommand{\thetable}{A\arabic{table}}
\begin{center}
\begin{longtable}{|llllll|}
\caption{Participant descriptives sorted to age. Range of motion data were not captured in three of the participants. A more extended version of this table, including body segment circumferences, can be found on the OSF repository (\href{https://osf.io/b4ru8/?view\_only=0b8e39fbf7924fbfb8df2ec666b6d986}{https://osf.io/b4ru8/?view\_only=0b8e39fbf7924fbfb8df2ec666b6d986}).} \label{tab:long} \\

\hline \multicolumn{1}{|c}{\textbf{Age}} & \multicolumn{1}{|c|}{\textbf{Sex}} & \multicolumn{1}{c|}{\textbf{Body weight}} & \multicolumn{1}{c|}{\textbf{Body height}} &  \multicolumn{1}{c|}{\textbf{L5S1 flexion}} &  \multicolumn{1}{c|}{\textbf{L5S1 extension}} \\ 
\multicolumn{1}{|c}{\emph{years}} & \multicolumn{1}{|c|}{\emph{Male/Female}} & \multicolumn{1}{c|}{\emph{kg}} & \multicolumn{1}{c|}{\emph{cm}} &  \multicolumn{1}{c|}{ \emph{degrees}} &  \multicolumn{1}{c|}{\emph{degrees}} \\
\endfirsthead
 \hline 
    25 & F & 59 & 171.5 & 38.6 & 25.1 \\
    31 & M & 70 & 180 & N/A & N/A \\
    31 & M & 80 & 181.5 & N/A & N/A \\
    32 & M & 78 & 192.5 & N/A & N/A \\
    34 & F & 62 & 174 & 29.9 & 22.1 \\
    36 & F & 61 & 161.5 & 32.8 & 40.8 \\
    38 & F & 72 & 172.5 & 52.2 & 28.3 \\
    40 & M & 77.5 & 181.5 & 54.2 & 23.5 \\
    43 & M & 77 & 180.5 & 53.3 & 13.8 \\
    48 & M & 100 & 194.5 & 56.9 & 24.8 \\
    48 & F & 63 & 188 & 53 & 15 \\
    52 & M & 85 & 186 & 59.8 & 26.2 \\
    52 & F & 88 & 169 & 29.7 & 13.2 \\
    53 & F & 81.5 & 161.5 & 9.7 & 25.9 \\
    54 & F & 83 & 168.5 & 27 & 18.3 \\
    57 & F & 69.5 & 167.5 & 49.6 & 16.3 \\
    59 & M & 79 & 187.5 & 34.3 & 18.6 \\
    60 & M & 90 & 194.5 & 38.1 & 13.6 \\ 
   \hline
\end{longtable}
\end{center}
\newpage 
\subsection{Postural differences between conditions}

\setcounter{figure}{0}    
\begin{figure}[h!]
    \centering
    \includegraphics[width = \textwidth]{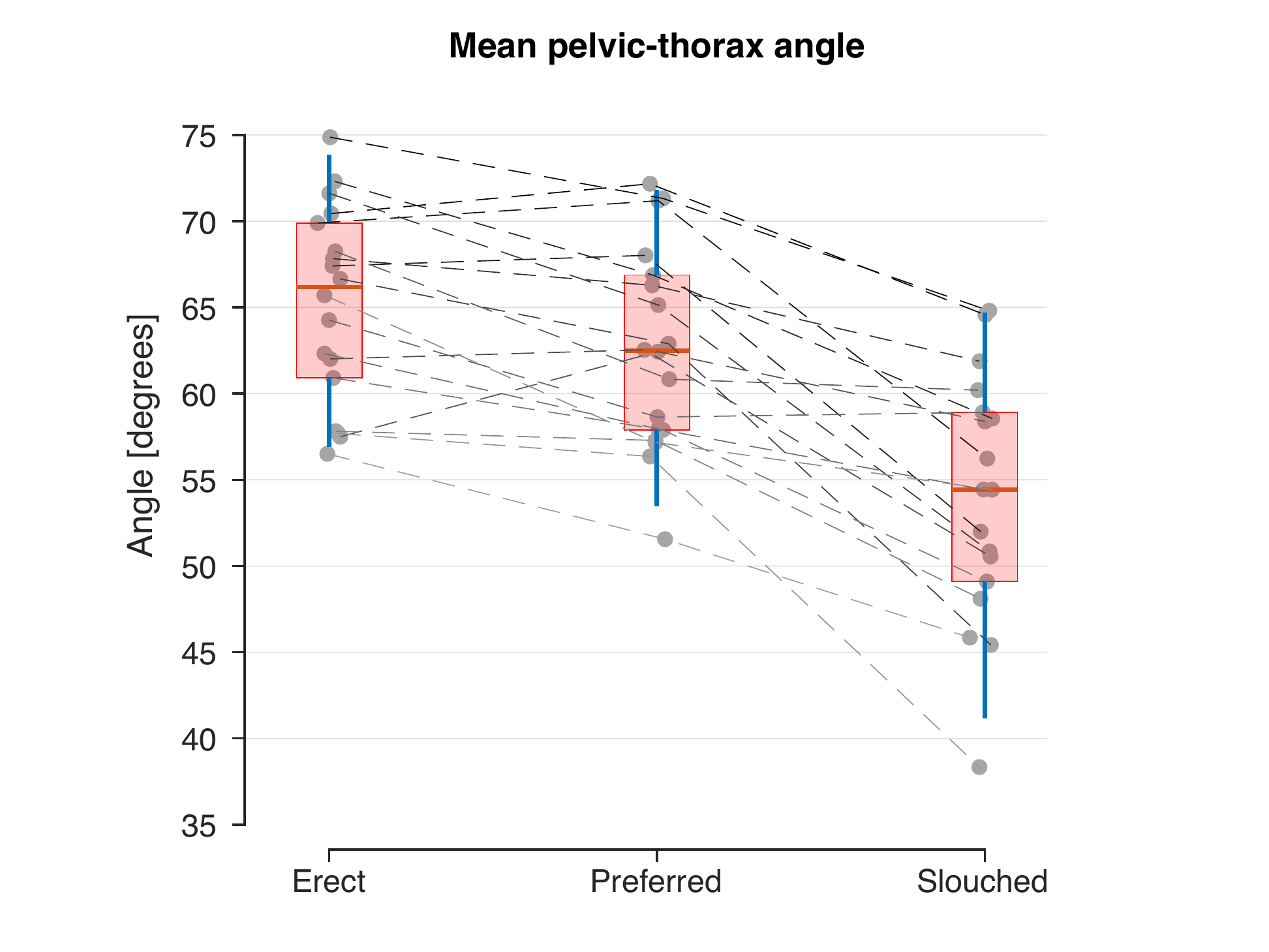}
    \caption{Effect of posture on pelvis and lumbar spine orientation. Box plots of the angle between the horizontal plane and the connecting line between the pelvis and thorax. Whiskers display the 5\textsuperscript{th} and 95\textsuperscript{th} percentile. Grey dots show the individual subject datapoints. To emphasise the within subject comparison, dotted lines are plotted to visualise the individual changes across conditions.}
    \label{sec:sample:conditions}
\end{figure}

\subsection{Ranges of Motion}
After the experiment the active ranges of motion of the entire spine (flexion/extension and lateral flexion) were recorded. As such, we can relate the sitting posture to the maximum flexibility of the spine. For instance, passive structures will increasingly contribute to postural stabilization when the lumbar flexion approximates the range of motion in slouched postures. 

We instructed the participants to stand straight and upright. Then, we asked participants to flex their trunks as far as possible until they felt their pelvis rotating. We instructed the participants to make a roll bending movement in a cranial-to-caudal order, so they first bent the cervical spine and further bent the spine until the pelvis rotated. This protocol was adapted from \cite{frey2020}.

\clearpage

\section{Seat pressure and center of pressure}
\label{sec:sample:appendix}
\renewcommand{\thefigure}{B.\arabic{figure}}

\noindent Seat pressure data were unevenly sampled and therefore resampled using linear interpolation.

\section*{Centre of pressure}
The center of pressure (CoP) location was derived, and the root-mean-square (rms) excursion was calculated (Figure B.1). 
CoP rms forward displacement was on average below 0.2 mm in all conditions and lateral displacement was also below 0.2 mm with fore-aft and vertical perturbations (Figure B.1). The lateral CoP displacement with lateral perturbations was significantly higher at 1.6 mm. The CoP rms analysis showed a main effect of excitation direction (F(2) = 104.48, p <0.001) and response direction (i.e., fore-aft and lateral; F(2) = 198.13, p <0.001). Furthermore, the excitation$\times$response direction interaction showed to modify the CoP rms values (F(2) = 588.48, p <0.001). A post hoc analysis, showed that CoP rms displacements were indeed significantly larger for the lateral excitation.

Transfer functions of the CoP motion relative to accelerations of the motion platform were calculated, with a similar approach as the transfer functions of the kinematics. The frequency domain analysis for the lateral CoP displacement showed substantial variation between participants with a moderate and varying coherence (Figure B.2).

\section*{Apparent mass}
For the vertical excitation, the apparent mass was calculated by computing the transfer functions of the total seat contact force (summation of pressure signals of individual sensors) relative to the vertical acceleration of the motion platform. Again, the transfer function was computed in a similar manner and with the same settings as for the translational and rotational frequency responses (see section \ref{text:freqresponse}). The apparent mass was very low with values around 15 kg (Figure B.3). Coherence varied strongly between participants. The low dynamic mass may be partially due to load transfer through the back support. However, a more likely explanation is that the pressure sensors underestimate dynamic loads calling for dynamic calibration \cite{Liu2018274} and verification measuring seat forces. 

\section*{Outlook}
Hence we conclude that seat pressure shows partially useful dynamic responses in lateral COP displacement with lateral seat motion and total force with vertical seat motion. Dynamic loads may well be underestimated which shall be taken into account when using pressure data to support modelling the human to seat interaction.

\setcounter{figure}{0}

\begin{figure}[p]
    \centering
    \includegraphics[width=\linewidth]{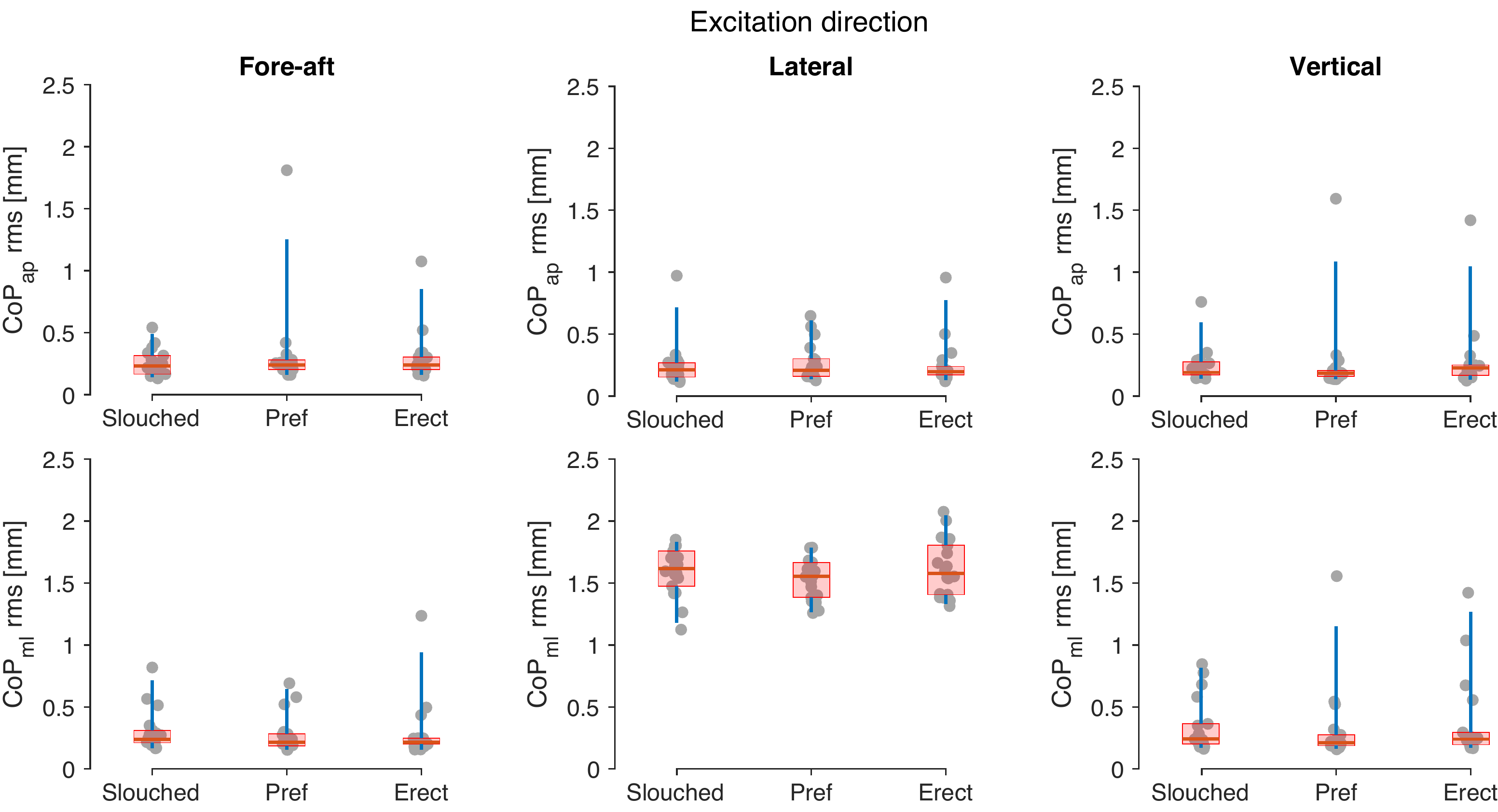}
    \caption{Centre of pressure (CoP) displacement. Upper panels show boxplots of the root mean square (rms) CoP displacements in anterior-posterior direction. Lower panels show the CoP rms displacements in lateral direction. The individual datapoints were overplotted as grey dots.\label{fig:cop}}
\end{figure}
\color{black}

\begin{figure}[!h]
    \centering
    \includegraphics[width =  1\linewidth]{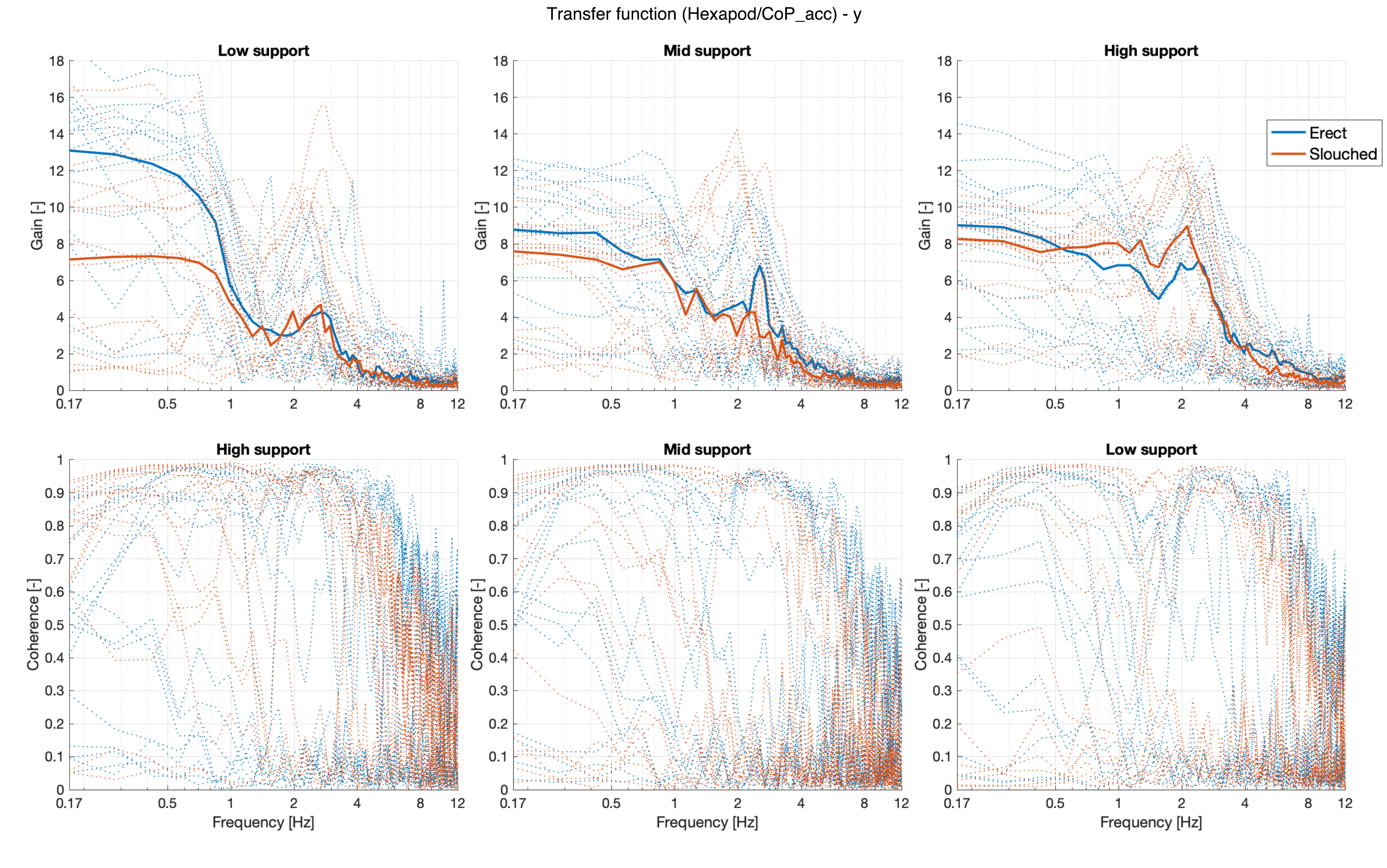}
    \caption{Transfer function from seat acceleration to seat centre of pressure (CoP) displacement in 
    lateral direction. See further Fig B.3.}
    \label{fig:tf_cop}
\end{figure}

\begin{figure}
    \centering
    \includegraphics[width=\textwidth]{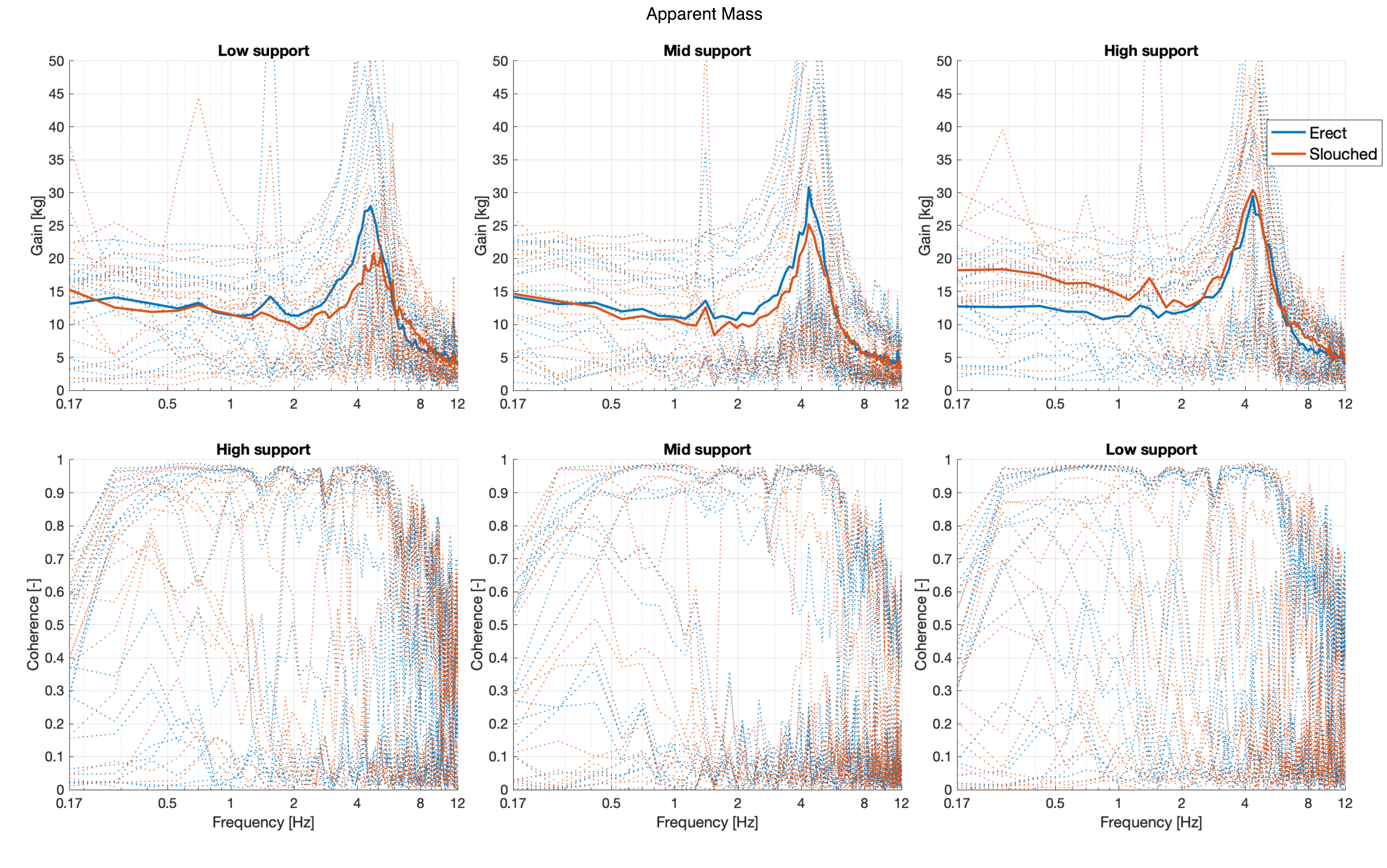}
    \caption{Apparent mass with vertical perturbations derived from the total seat contact force (summation of pressure signals of individual sensors) relative to the vertical acceleration of the motion platform. Transfer function gain (top panels) and coherence (bottom panels). The dotted lines depict the individual results, whereas the superimposed continuous lines show the average across participants. Only the transfer functions under the erect (blue) and slouched (red) conditions are shown. Software to import and visualize XSENSOR seat pressure data can be obtained from \href{https://github.com/nickkluft/XSensorPressureMat}{https://github.com/nickkluft/XSensorPressureMat}.}
    \label{fig:tf_appmass}
\end{figure}

\newpage
\section{EMG analysis}
\label{sec:sample:appendix}

Muscular activity was evaluated recording surface EMG from 8 muscles from 4 subjects. Gel-based electrodes were applied with an interelectrode distance of 2 cm and raw EMG was recorded at 2048 Hz. Electrodes were placed in accordance with the Seniam guidelines. Prior to electrode placement, the skin was shaved and cleaned with alcohol. In the neck we measured the left and right capitis semispinalis, the left and right upper trapezius, and the left and right sternocleidomastoid. In the lumbar area we measured the left and right multifidus at the level of the spinous process of L5. In pilots we also measured the rectus abdominis but no coherent activity was found.

We first analysed EMG with high back support with the preferred posture and focused on the anterior-posterior motion case. With high support we expected clear and consistent activity in the neck muscles. We also explored results with low back support where more lumbar activity was expected. 

The raw EMG was bandpass filtered (65 – 800 Hz : see below) and rectified. Gain, phase and coherence were calculated with respect to the applied platform motion. Coherences for all muscles were investigated after the application of both 1st and 6th order bidirectional Butterworth filters. Filter parameters were tuned to achieve a high coherence from 1-6 Hz being the range where coherence was highest in the neck muscles. Application of a 6th order Butterworth filter led to highest coherence. Thus, the 6th order Butterworth filter was chosen for analysis.

A Hamming window was applied using 10 segments with 50\% overlap over the selected 65 seconds of data. The applied bandpass filter frequency range was tuned to achieve a good coherence. The lower cutoff frequency (i.e. 65 Hz) was found to strongly affect the coherence. Therefore, after the application of various cutoff frequencies in the range of 5 – 205 Hz, coherences for all muscles were explored. Eventually, a lower cutoff frequency of 65 Hz was chosen as this cutoff frequency showed best results. On the contrary, shifting the upper cutoff frequency had limited effect. After tuning these parameters, coherence improved and gain and phase showed increased consistency between left and right muscles. 

We varied the number of segments between 2 – 20 and determined the significance threshold for coherence as this demonstrates whether the coherence differs from zero. Lower segment numbers improved the coherence but also increased the significance threshold as the threshold depends on the number of segments. Hence, 10 segments were selected for further analysis.

EMG and kinematics disclosed some voluntary motions including changes in the head orientation. Attempts to eliminate such events from the frequency domain analysis were not successful possibly due to the irregular and limited duration of remaining time segments.

After filtering and tuning of the filter parameters, normalization towards the rms EMG was performed and left and right muscles were averaged. Furthermore, results obtained from trials in which subjects had to sit in an erect or preferred posture with the back seat set in the mid or highest position, were also averaged. As a consequence of these averaging methods, coherence improved for the neck muscles but not for the lumbar muscle (Multifidus). 

In the end, significant coherence in the range of 1 – 6 Hz was obtained for Sternocleidomastoid and Trapezius, coherence was close to significant for Semispinalis and remained insignificant for Multifidus which was the only lumbar muscle studied (\cref{EMG},\cref{EMG2}). 

These results may be usable to validate models of neck stabilization, but these EMG data seem not informative of lumbar stabilization. Here we need to consider that muscular co-contraction and seat back support will also contribute to trunk stabilization in car occupants. However, pilot measurements showed that with higher motion amplitudes and longer exposure, EMG coherence for lumbar muscles enhanced. In future studies we may explore transient responses such as strong braking and steering. We may also explore usage of more electrodes or electrode arrays to enhance sensitivity and specificity.

\setcounter{figure}{0}    
\renewcommand{\thefigure}{C.\arabic{figure}}
\begin{figure}[h!]
\centering
\includegraphics[trim = 100 30 100 20,width = 1\textwidth]{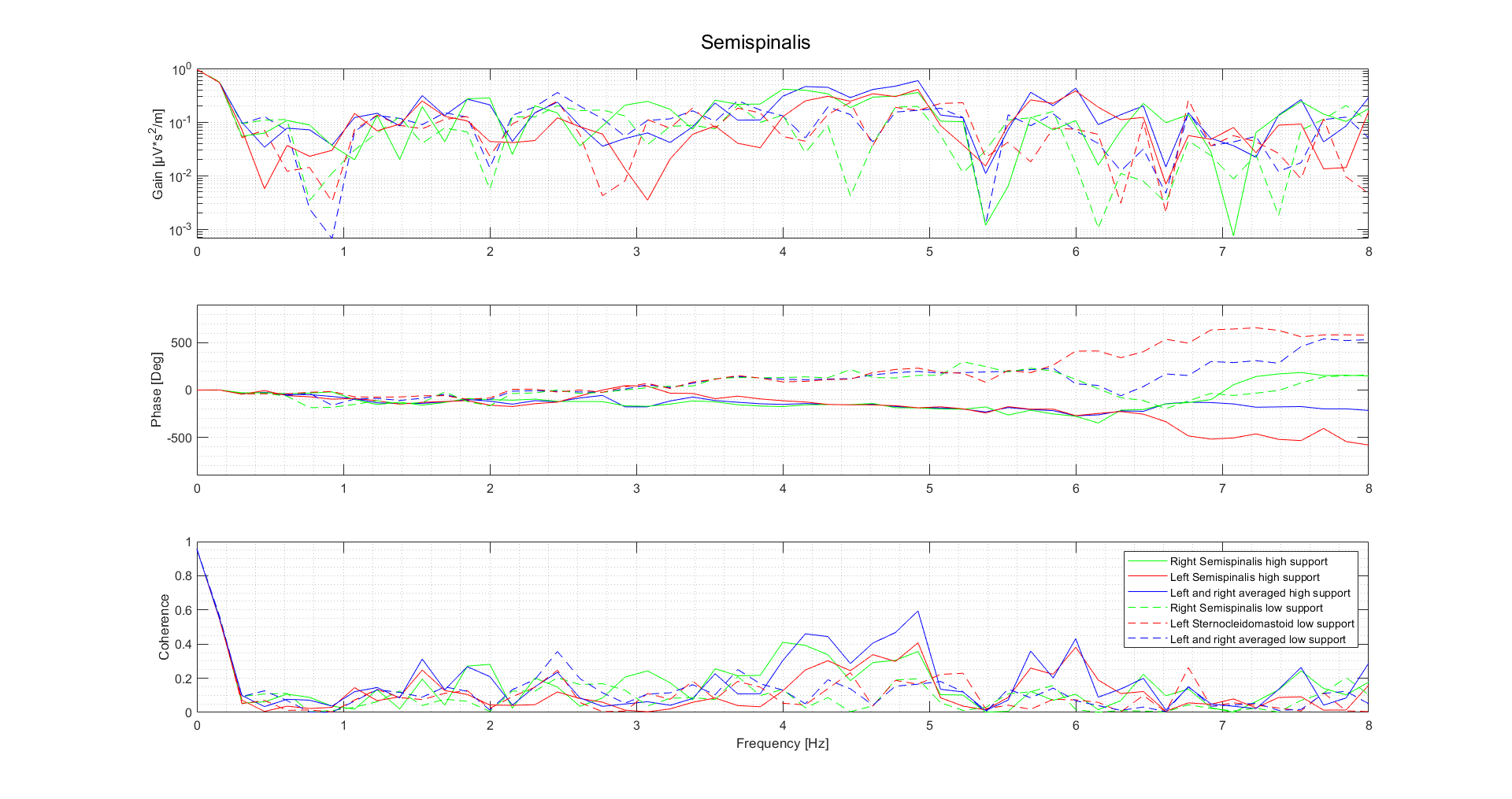}
\includegraphics[trim = 100 30 100 20,width = 1\textwidth]{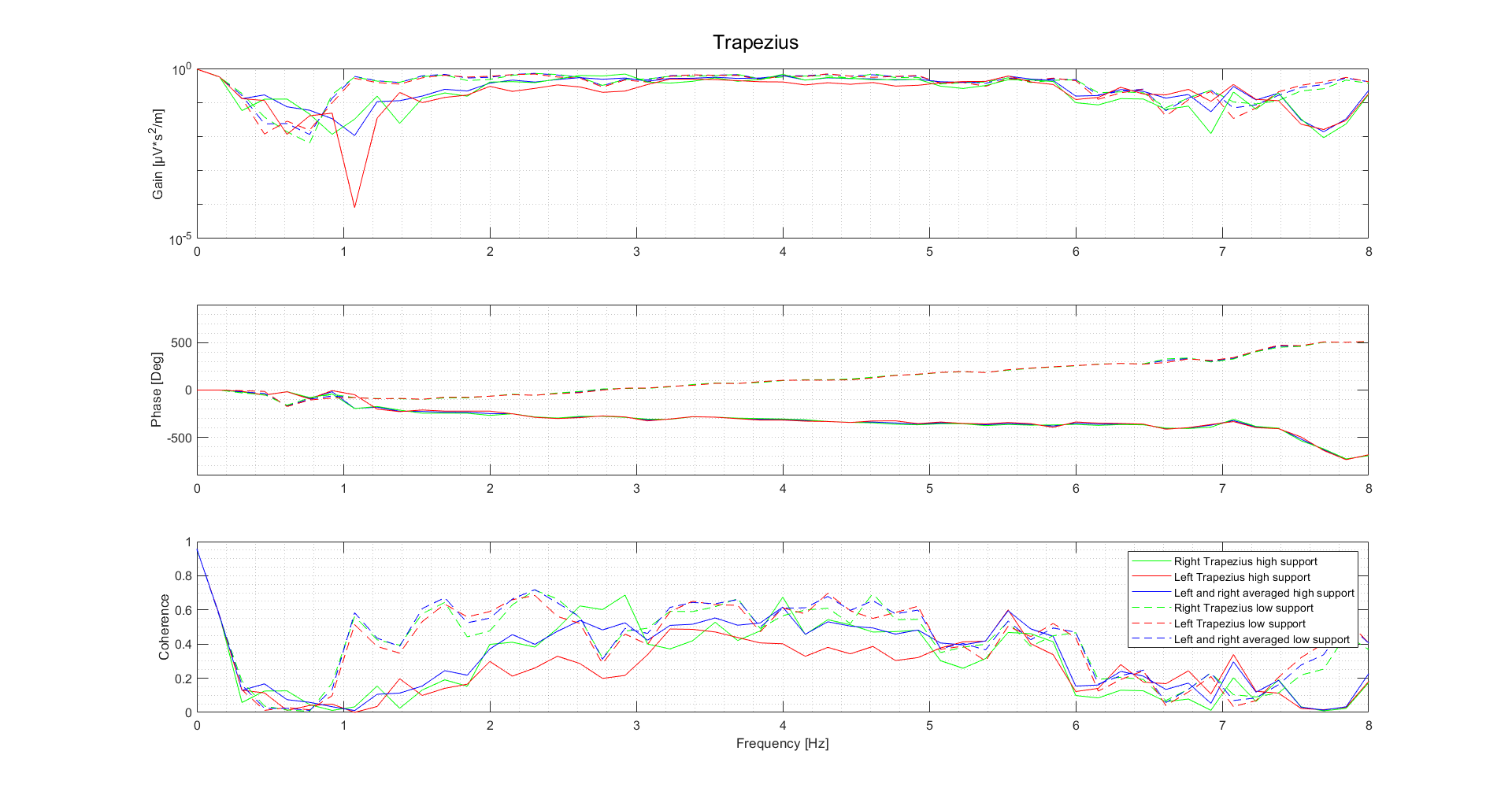}

\caption{Semispinalis and Trapezius (neck) EMG responses to platform acceleration: Gain, phase and coherence \label{EMG}}
\end{figure}

\begin{figure}[!h]
\centering
\includegraphics[trim = 100 30 100 20, width = 1\textwidth]{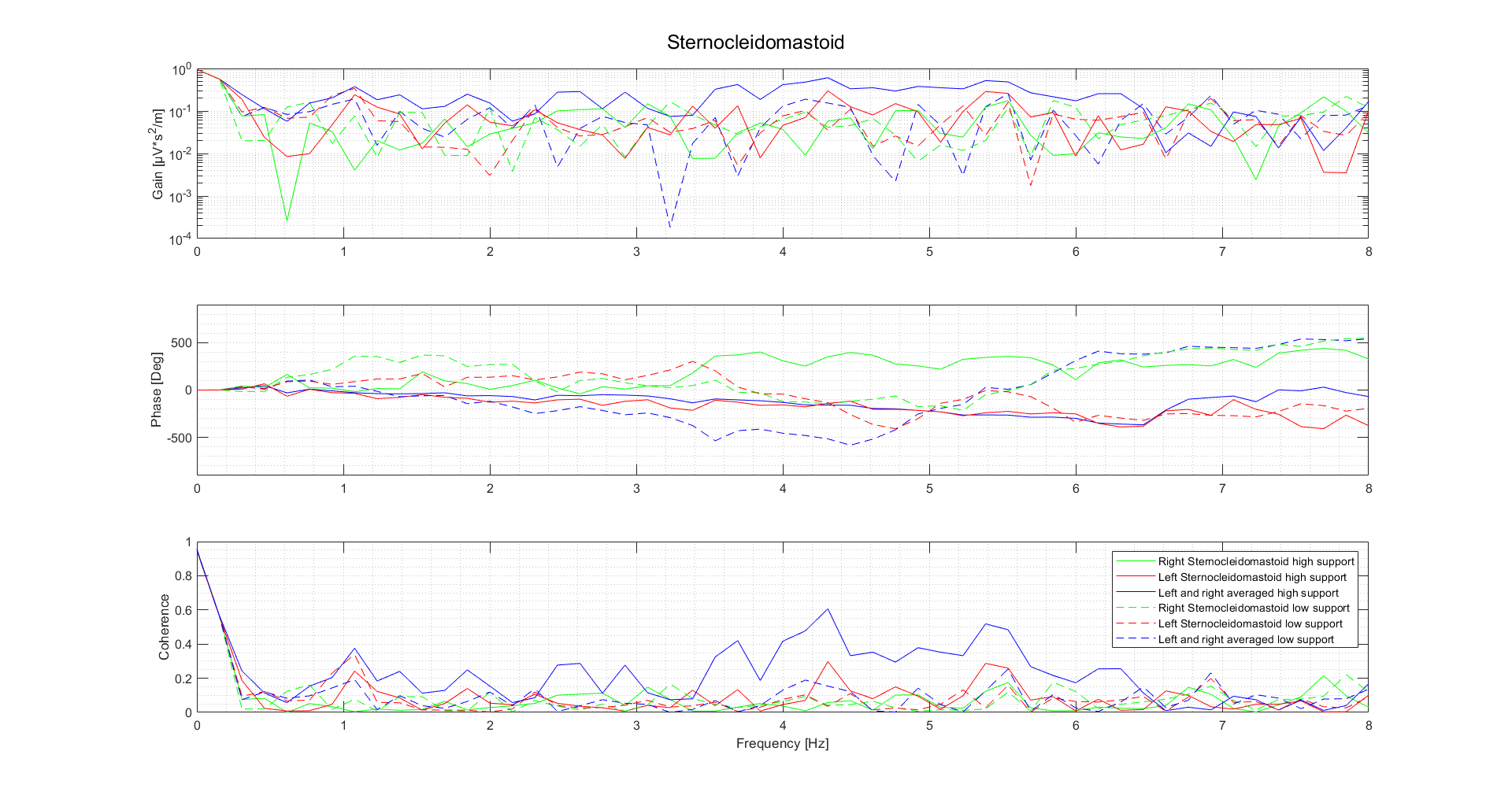}
\includegraphics[trim = 100 30 100 20,width = 1\textwidth]{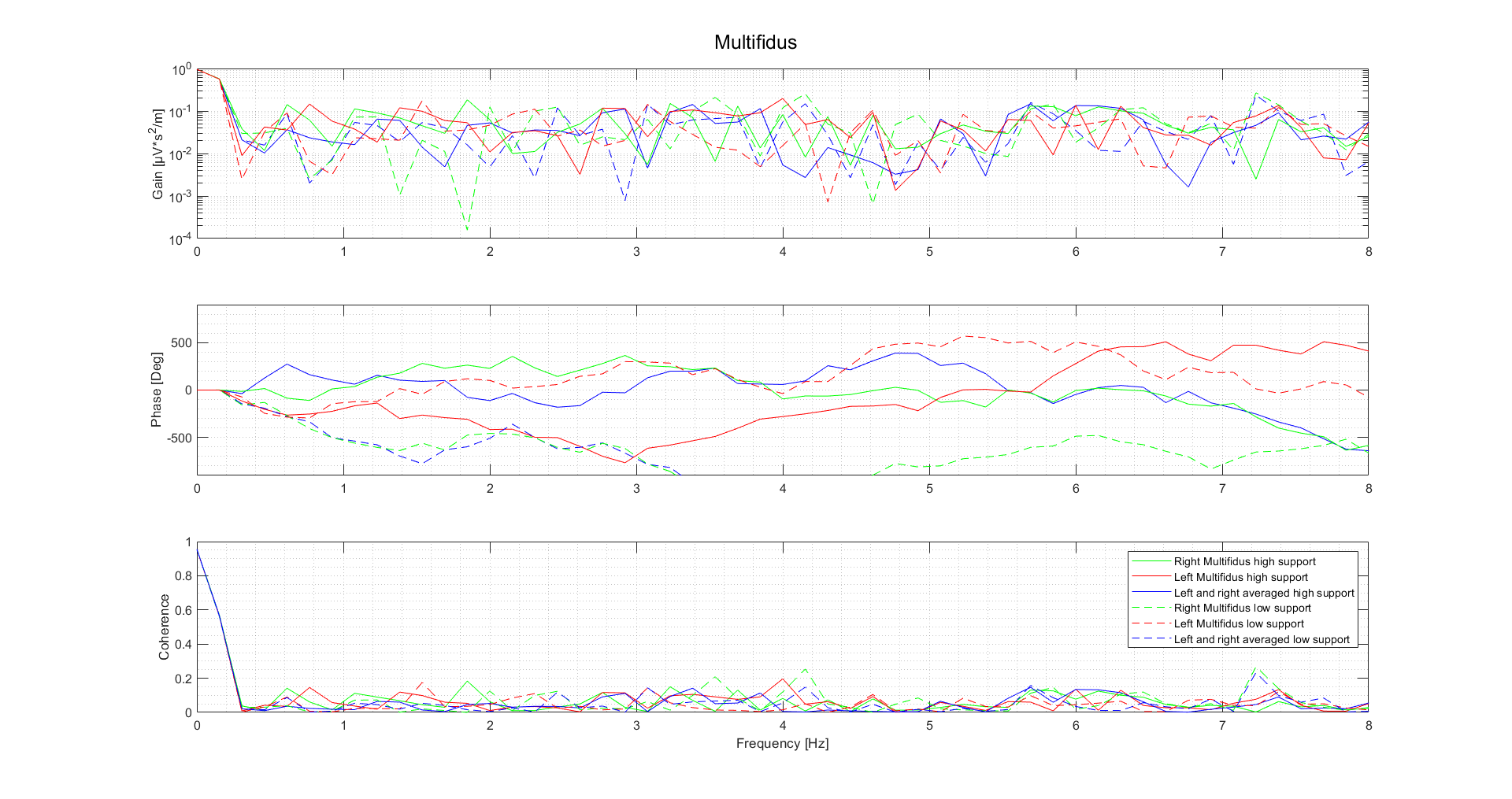}
\caption{Sternocleidomastoid (neck) and Multifidus (lumbar) EMG responses to platform acceleration: Gain, phase and coherence \label{EMG2}}
\end{figure}

\section{Frequency responses for most conditions and all participants}
\label{sec:sample:appendix}


This appendix contains body responses (head, trunk and pelvis) to car mock-up perturbations (see the main paper for more details). 
Following figures provide frequency domain analysis of recorded kinematic data in terms of Gain, Phase and Coherence.
Dotted lines indicate each individual subject response, blue solid lines represent the median and the solid black lines are 25\textsuperscript{th} and 75\textsuperscript{th} percentiles.

\vspace{10mm}
\noindent Data for 9 conditions are shown: 
\begin{itemize}
    \item 3 back support heights (Low, Medium and High) in 2 posture conditions (Erect, Slouched) plus 
    \item 3 special cases (Eyes Closed, Head Down and Low Amplitude).
\end{itemize}

\noindent For each condition 7 figures show:
\begin{itemize}
\item Figure D.x.1: 3*3 (x: paragraph number) only showing the main translational responses for the perturbation in that direction, being the surge response (X) for the perturbation in surge (left column), the lateral response to lateral perturbations (mid column), and the vertical response to vertical perturbations (right column). 
\par Following figures shows all 3D translational and rotational responses
\item Figure D.x.2: Translational responses to fore-aft perturbations.
\item Figure D.x.3: Rotational responses to fore-aft perturbations.
\item Figure D.x.4: Translational responses to lateral perturbations.
\item Figure D.x.5: Rotational responses to lateral perturbations.
\item Figure D.x.6: Translational responses to vertical perturbations.
\item Figure D.x.7: Rotational responses to vertical perturbations.
\end{itemize}
\includepdf[pages=-]{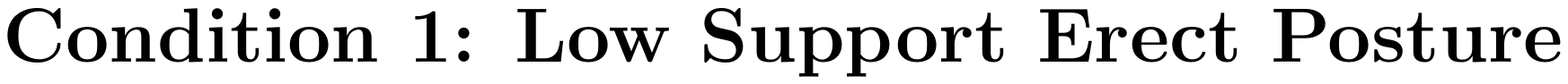}
\section{Full data set}
\label{sec:sample:appendix}
Supplied as matlab files in the OSF repository 








\end{document}